%% file: sdss2222lens_ms.tex
\documentclass[apj]{emulateapj}
\usepackage{amsmath}
\usepackage{natbib}
\usepackage{color}
\usepackage{url}

\newcommand{\lenstool}{{\tt{Lenstool}}}

\newcommand{\Lenstool}{{\tt{Lenstool}}}

\newcommand{\clustername}{SDSS\,J2222$+$2745}
\newcommand{\zcluster}{0.49}
\newcommand{\zQSO}{2.805}
\newcommand{\zsource}{2.805}
\newcommand{\zarcA}{2.3}
\newcommand{\zarcB}{4.56}
\newcommand{\hst}{{\it HST}}

\newcommand{\Swift}{{\it Swift}}
\newcommand{\Lya}{Ly$\alpha$}
\newcommand{\kms}{km s$^{-1}$}
\newcommand{\msun}{M$_{\odot}$}
\newcommand{\tac}{$\tau_{\rm AC}$}
\newcommand{\tab}{$\tau_{\rm AB}$}
\newcommand{\tad}{$\tau_{\rm AD}$}
\newcommand{\tae}{$\tau_{\rm AE}$}
\newcommand{\taf}{$\tau_{\rm AF}$}
\newcommand{\dahleA}{Dahle et al. (2013)}
\newcommand{\dahleB}{Dahle et al. (2015)}

\newcommand{\implanerms}{$0\farcs 16$}
\newcommand{\qsoAdist}{34}

\newcommand{\timeAB}{$ 47 \pm{ 20}$}
\newcommand{\timeAC}{$-726 \pm{294}$}

\newcommand{\timeABobs}{$47.7 \pm{6.0}$}
\newcommand{\timeACobs}{$-722 \pm{ 24}$}

\newcommand{\timeADpred}{$502 \pm{ 68}$}
\newcommand{\timeAEpred}{$611 \pm{ 75}$}
\newcommand{\timeAFpred}{$415 \pm{ 72}$}
\newcommand{\timeDEpred}{$102 \pm{23}$}
\newcommand{\timeDFpred}{$-85 \pm{39}$}

\newcommand{\magA}{$14.5 \pm 2.7$}
\newcommand{\magB}{$10.8 \pm 4.3$}
\newcommand{\magC}{$6.7 \pm 1.0$}
\newcommand{\magD}{$1.43 \pm 0.75$}
\newcommand{\magE}{$0.76 \pm 0.39$}
\newcommand{\magF}{$0.95 \pm 0.65$}

\newcommand{\monehundred}{$M_{(<100{\rm kpc})}$ = $0.55  \pm 0.03 \times 10^{14}$~\msun}
\newcommand{\mtwohundred}{$M_{(<200{\rm kpc})}$ = $1.15 \pm 0.1 \times 10^{14}$~\msun}
\newcommand{\mfivehundred}{$M_{(<500{\rm kpc})}$ = $2.50 \pm 0.24 \times 10^{14}$~\msun}

\newcommand{\Mdyn}{$M_{200,{\rm dynamical}} = 3.9 ^{+3.6}_{-2.2} \times 10^{14}$~\msun}
\newcommand{\vdisp}{$\sigma_{v} = 657 \pm 166$ \kms}

\newcommand{\Mxraylowhi}{$M_{500,{\rm x}} = [0.94-1.2] \times  10^{14}$~\msun}

\input{fixed_parameters.tex}

\input{parameters.tex}

\shorttitle{The lens model of \clustername}
\slugcomment{ApJ in prep: draft date \today}
\shortauthors{Sharon et al.}

\begin{document}
\title{Lens model and time delay predictions for the sextuply lensed
  quasar \clustername \altaffilmark{*}}
\altaffiltext{*}{Based on observations made with the NASA/ESA {\it Hubble
  Space Telescope}, obtained at the Space Telescope Science Institute,
  which is operated by the Association of Universities for Research in
  Astronomy, Inc., under NASA contract NAS 5-26555. These observations
  are associated with program GO-13337.} 

\author{Keren Sharon$^1$,
Matthew B. Bayliss$^{2,3,4}$,
H{\aa}kon Dahle$^5$,
Michael K. Florian$^6$,
Michael D. Gladders$^{6,7}$,
Traci L.  Johnson$^1$,
Rachel Paterno-Mahler$^1$,
Jane R. Rigby$^8$,
Katherine E. Whitaker$^{9\dagger}$, 
\and Eva Wuyts$^{10}$}

\footnotetext[$\dagger$]{Hubble Fellow}
\email{kerens@umich.edu} 

\affiliation{
\altaffiltext{1}{Department of Astronomy, University of Michigan, 1085 S. University Ave, Ann Arbor, MI 48109, USA}
\altaffiltext{2}{Colby College, 5800 Mayflower Hill, Waterville, 04901, Maine, USA}
\altaffiltext{3}{Harvard-Smithsonian Center for Astrophysics, 60  Garden Street, Cambridge, MA 02138, USA}
\altaffiltext{4}{Department of Physics, Harvard University, 17 Oxford St., Cambridge, MA 02138}
\altaffiltext{5}{Institute of Theoretical Astrophysics, University of  Oslo,  P. O. Box 1029, Blindern, N-0315 Oslo, Norway }
\altaffiltext{6}{Department of Astronomy and Astrophysics, University of Chicago, 5640 South Ellis Avenue, Chicago, IL 60637, USA}
\altaffiltext{7}{Kavli Institute for Cosmological Physics, University of Chicago, 5640 South Ellis Avenue, Chicago, IL 60637, USA.}
\altaffiltext{8}{Astrophysics Science Division, Goddard Space Flight Center, 8800 Greenbelt Rd., Greenbelt, MD 20771}
\altaffiltext{9}{Department of Astronomy, University of Massachusetts--Amherst, Amherst, MA 01003, USA}
\altaffiltext{10}{Max-Planck-Institut f{\"u}r extraterrestrische Physik, Giessenbachstr. 1, D-85741 Garching, Germany }
}

\begin{abstract}
\clustername\ is a galaxy cluster at $z=\zcluster$, strongly lensing a
quasar at $z=\zQSO$ into six widely separated images. 
In recent \hst\ imaging of the field, we identify additional multiply lensed
galaxies, and confirm the sixth quasar image that was identified by
Dahle et al. (2013). We used the Gemini North telescope to measure a spectroscopic
redshift of $z=\zarcB$ of one of the secondary lensed galaxies. These data are used
to refine the lens model of \clustername, compute the time delay
and magnifications of the lensed quasar images, and reconstruct the
source image of the quasar host and a second lensed galaxy at
$z=$\zarcA. This second galaxy also appears in absorption in our Gemini spectra of
the lensed quasar, at a projected distance of \qsoAdist\ kpc.  
Our model is in
agreement with the recent time delay measurements of Dahle et
al. (2015), who found \tab=\timeABobs\ days and \tac=\timeACobs\ days. 
We use the observed time delays to further constrain the model, and
find that the model-predicted time delays of the three faint images of the quasar
are \tad=\timeADpred\ days, \tae=\timeAEpred\ days, and
\taf=\timeAFpred\ days. We have initiated a
follow-up campaign to measure these time delays with Gemini
North. Finally, we present initial results from an X-ray monitoring program with \Swift,
indicating the presence of hard X-ray emission from the lensed
quasar, as well as
extended X-ray emission from the cluster itself, which is consistent
with the lensing mass measurement and the cluster velocity
dispersion. 
\end{abstract}


\section{Introduction}
The rare chance alignment of a quasar behind a strong-lensing cluster
provides unique opportunities for studies of different astrophysical
objects. 
Through careful lens modeling, these systems can probe the mass
distribution of the
foreground lens; the high magnification enhances our ability to study
the background quasar, and galaxies between us and the quasar 
 can be seen in absorption along multiple lines of sight in the light of the background
quasar. 
Lensing configurations that involve a quasar lensed by a single
massive galaxy are more
common; however,  the lensing magnification of a single galaxy is
typically significantly lower than in the galaxy cluster case. 
Unique to the cluster-lensed quasar
configurations, the multiple images of the lensed quasar have large
separations ($14\farcs6 - 22\farcs5$; Inada et al. 2003, 2006; Dahle et
al. 2013)  and high magnifications; the lensed active nucleus is
point-like, providing accurate positional constraints, and is
variable –- enabling measurements of the time delay between images of the
same source. The high tangential magnification stretches the host galaxy of the
quasar into a giant arc, thus resolving it from the light of
the active nucleus, which usually dominates in a high-redshift quasar.  

To date, only three cases of high-redshift
quasars strongly-lensed
by a galaxy cluster  are published: SDSS J1004+4112 (Inada et al. 2003), SDSS
J1029+2623 (Inada et al. 2006), and \clustername\ (Dahle et
al. 2013). 

\clustername\ was discovered as part of the Sloan Giant Arcs Survey
(SGAS; Gladders et al. in prep, Bayliss et al. 2011a,b; Hennawi et
al. 2008; Sharon et al. 2014). SGAS is a systematic survey of highly magnified
lensed galaxies, also refered to as ``giant arcs,'' in the imaging data of the Sloan
Digital Sky Survey (SDSS, York et al. 2000). The lensing identification
process starts with optical selection of galaxy clusters from the SDSS photometry
catalogs, using the cluster red sequence algorithm of Gladders \& Yee
(2000). Sections of the imaging data around each cluster were then
retrieved and processed to generate color images, with scaling
parameters selected to optimize the visibility of possible lensing
features. The images were visually inspected and ranked for lensing evidence by
several observers in a process that enables a calculation of the
selection statistics (the process will be described in full in
Gladders et al., in preparation). All candidates were followed up for
confirmation, and the survey purity and completeness were quantified. Bayliss et
al. (2011a,b) give the results of the initial spectroscopic followup campaign,
and measure the redshift distribution of the lensed
galaxies. 

\clustername\ was detected in the SGAS search in SDSS Data
Release 8 (Aihara et al. 2011) owing to a prominent giant arc that
appears $8\farcs5$ south of the brightest cluster galaxy. 
A further investigation of the field revealed the
multiply-imaged lensed quasar. 
The field was followed up by Dahle et al. (2013) 
using the Mosaic Camera (MOSCA) and the Andalucia Faint Object
Spectrograph and Camera (ALFOSC) at the 2.56 m Nordic Optical Telescope
(NOT).

\begin{figure*}
\centering
\includegraphics[scale=0.5]{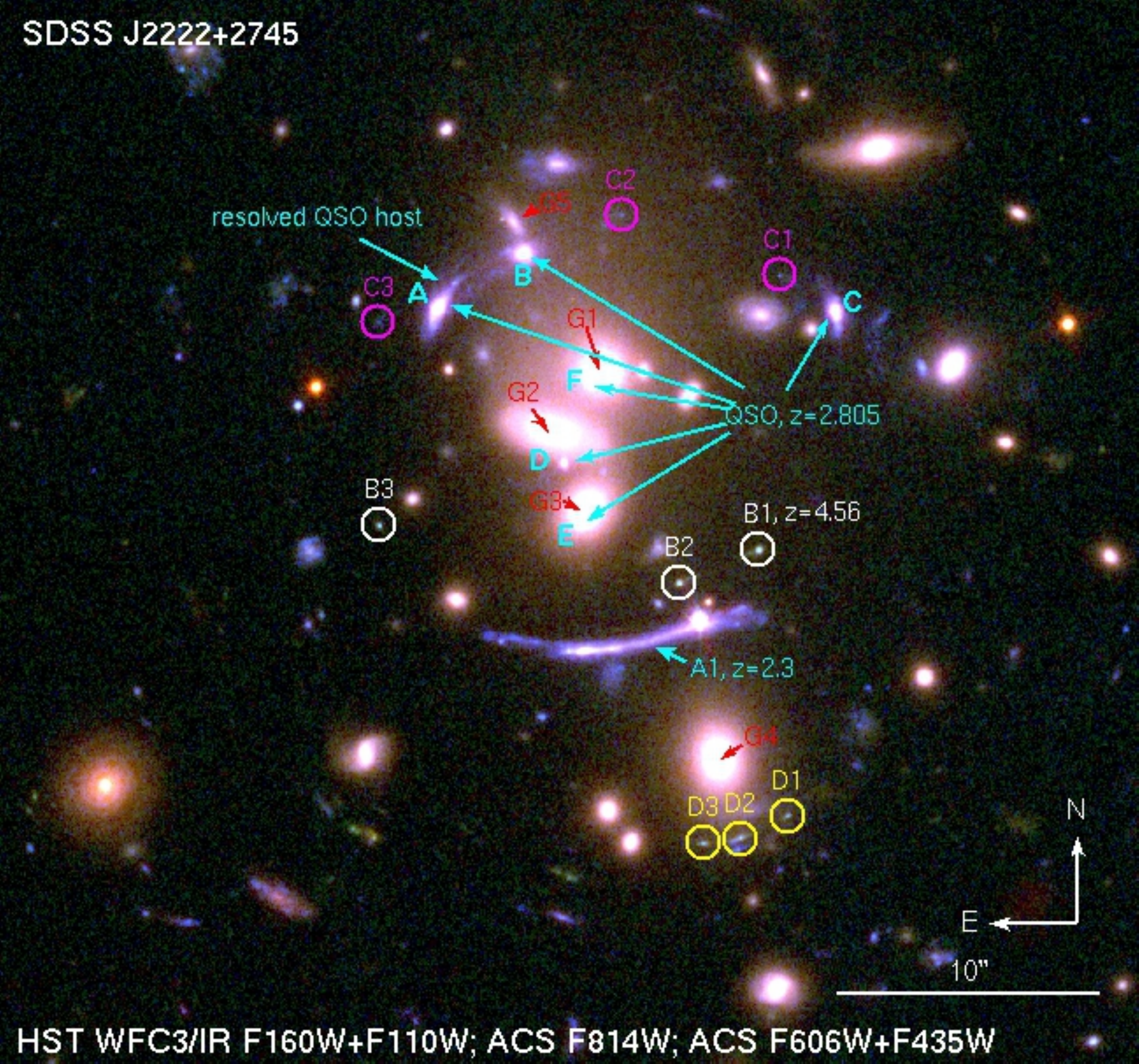}
\caption{Color composite image of \clustername\ from our \hst\ program
  GO-13337 (PI: Sharon) in WFC3/IR
  F160W+F110W (red), ACS F814W (green), ACS F606W+F435W (blue). These
  data confirm the sixth quasar image (F) that was identified by
  \dahleA\ as tentative.  The
  six images of the lensed quasar at $z=\zQSO$ and the previously
  identified giant arc A1 at $z=\zarcA$ are labeled in cyan; newly discovered
secure multiply-imaged galaxies are labeled in white (B1, B2, B3 at
$z=\zarcB$, see \S~\ref{s.spec}), yellow (D1, D2, D3) and magenta (C1, C2,
C3). Other possible arc candidates are not labeled. Note that the
point source that is seen embedded in the A1 arc is a foreground white
dwarf star (Dahle et al. 2013). }
\label{fig.hst}
\end{figure*}

We have recently obtained \hst\ imaging data of this target
(Figure~\ref{fig.hst}; Section~\ref{s.data}).
As can be seen in Figure~\ref{fig.hst}, a background quasar is lensed
by \clustername, forming six images around the core of a galaxy
cluster at $z=$\zcluster. Three bright images appear north
of the cluster core (labeled A, B, C; our labeling scheme follows Dahle
et al. 2015), and three faint images (D, E, F) can be seen near
the central cluster galaxies (G2, G3, G1, respectively).
The cluster also lenses other background galaxies,
the most prominent of which is seen as a blue arc south of the cluster core
(labeled A1 in Figure~\ref{fig.hst}). 
Dahle et al. (2013) reported on the discovery of \clustername, 
confirmed the lensing interpretation,
presented spectroscopic identification of the lensed quasar, spectroscopic
confirmation of the six lensed images of the quasar, and measured its
redshift to be $z=2.82$. 
In addition, we measured the spectroscopic redshifts of several
cluster member galaxies,
 and of the lensed galaxy A1 at $z=$\zarcA. 
Stark et al. (2013) also measure the spectra of the quasar, $z=2.807$
and of galaxy A1.
Interestingly, the spectrum of the quasar shows strong
\Lya\ absorption at the redshift of the foreground lensed galaxy, as
well as Si~II $\lambda$1526 and CIV~$\lambda$1549  (Stark et al. 2013), indicating the
presence of neutral hydrogen and metals associated with gas
surrounding the galaxy. Stark et al. (2013) estimated that the projected
distance between the quasar image A and the interloper galaxy A1 is $\sim
50$ kpc. We refine this estimate in Section~\ref{s.absorber}. 

Following the discovery of \clustername, we have initiated an imaging
monitoring program with the NOT to measure the time delays between the
images of the quasar. The results from the first three years of ongoing
photometric monitoring with the NOT and the first season of Gemini monitoring
are presented in Dahle et al. (2015). 
The light curves of the brighter three images of \clustername\ are
measured from an analysis of 42 distinct epochs, resulting in time delays
of \tab=\timeABobs\ days, and \tac=\timeACobs\ days. A
robust measurement of the time delays of images D, E, and F requires
deeper observations; a monitoring campaign with Gemini was initiated
in 2015 (GN-2016A-Q-28; PI: Gladders) for this purpose. 

This paper is structured as follows. In Section~\ref{s.data} we
describe the \hst\ imaging data of \clustername, Gemini spectroscopy, and \Swift\ X-ray
observations. We present a new strong lensing analysis based on the
new data in Section~\ref{s.lensing}. In Section~\ref{s.results}, we
present and discuss the predicted time delays, cluster
mass, lensing magnification, source reconstruction, and absorbing
systems. 
We conclude with future work in Section~\ref{s.conclusions}.
Throughout this paper, we assume a flat cosmology with $\Omega_{\Lambda}
= 0.7$, $\Omega_{m} =0.3$, and $H_0 = 70$  km s$^{-1}$
Mpc$^{-1}$. In this cosmology, $1\arcsec$ corresponds to 6.0384 kpc at
the cluster redshift, $z=$\zcluster. Magnitudes are reported in the AB system.

\section{Data}\label{s.data}
\subsection{\hst\ Imaging}\label{s.hst}
\begin{figure}
\centering
\includegraphics[scale=0.21]{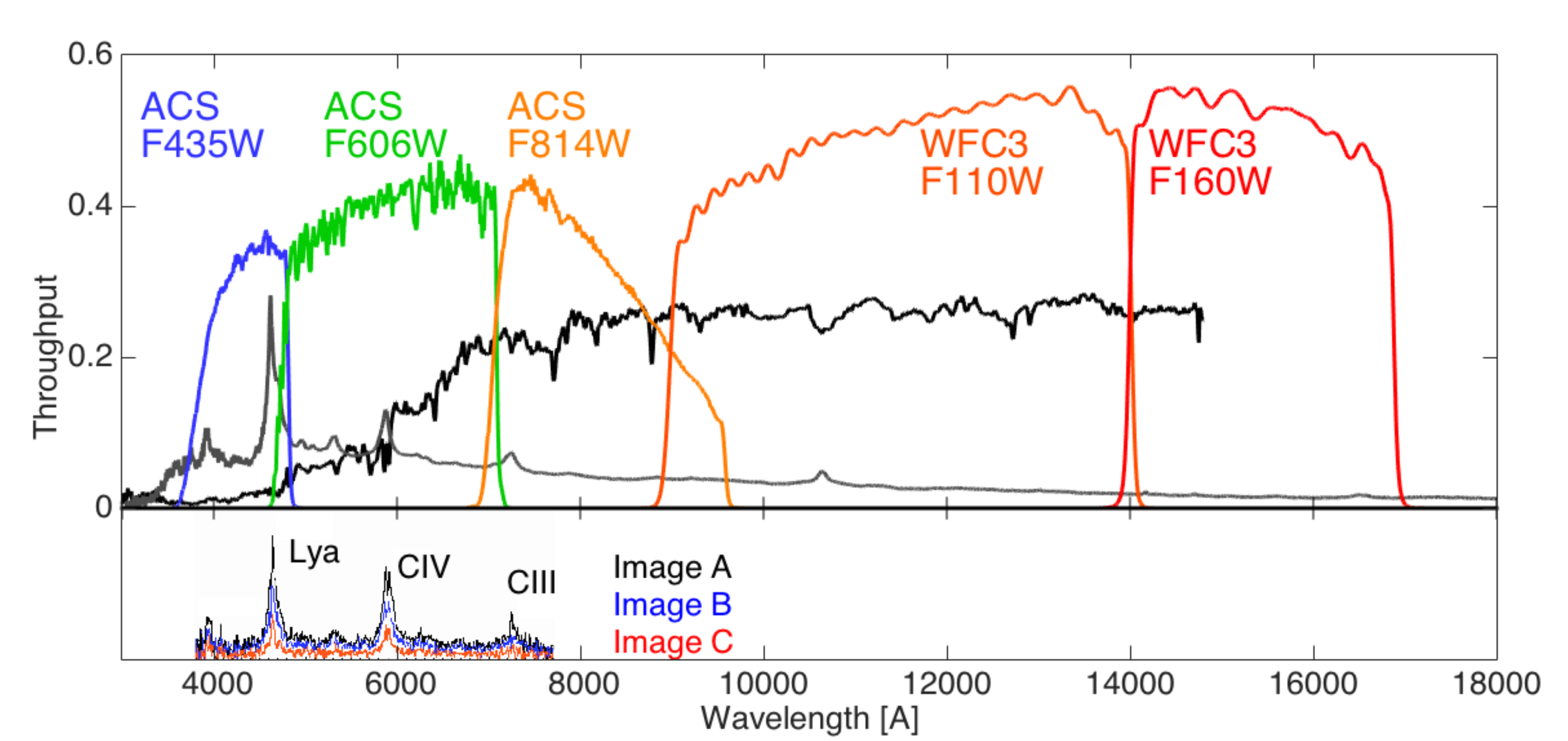}
\caption{The response function of the five filters that were used in
  the \hst\ observation (GO-13003; PI: Sharon) overplotted on a
  template spectrum of an elliptical galaxy, redshifted to the cluster redshift,
  $z=\zcluster$ (black), and a template quasar spectrum redshifted to the quasar
  redshift, $z=\zQSO$ (gray). The filters sample the spectral energy
  distribution of cluster galaxies as well as the quasar and its host,
  and provide good contrast between these sources. The broad
  wavelength coverage supports SED fitting and identification of
  secondary arcs. {\it Bottom}: Spectra of images A, B, and C of the
  quasar from Dahle et al. (2013).}
\label{fig.filters}
\end{figure}

\clustername\ was observed by \hst\ Cycle~21 program GO-13337 (PI:
Sharon) with WFC3 F160W for 1311~s and
F110W for 1211~s on 2014 Aug 10, and with ACS F435W, F606W, and F814W
for 4944~s each on 2014 Oct 10-11 
\footnote{The A, B, and C quasar images showed very little photometric
variation in the interval between the ACS and WFC3 observations: On 2014 Aug 5.06,  g(A)=21.61; g(B)=21.92; g(C)=21.95. On
2014 Oct 14.98, g(A)=21.62; g(B)=21.93; g(C)=21.92. All numbers are
from the ALFOSC/NOT monitoring reported in Dahle et al. (2015).}.  
The filters were carefully selected to provide
the best sensitivity to the different sources in the field. The bluest
filter, F435W, is sensitive to emission from the quasar and its host
and gives high contrast between the quasar and the cluster
galaxies, as can be seen in Figure~\ref{fig.filters}. At $z$=\zcluster\ most of the light from typical elliptical galaxies is
redshifted to wavelengths longer than the ACS/F435W response curve,
and we expect to see little emission in this band from the
cluster galaxies. ACS/F606W and ACS/F814W give good sampling of the
 spectral energy distribution of typical early-type cluster galaxies; and the reddest filters help
detect lensed high-$z$ dropout galaxies and provide a long wavelength
baseline for galaxy colors and SED fitting. 

Each of the ACS images was taken over two orbits, with three
gap-crossing sub-pixel 
dither positions in each orbit (a total of six sub-exposures) for better sampling of the point spread
function, removal of cosmic rays, hot or bad pixels, and to cover the chip
gaps. 
A half field-of-view offset was implemented between the two orbits of
observation in each filter. Since the strong lensing regime is small
enough to fit within
one ACS chip, this design ensures that the center of the field is imaged to the full depth of
two orbits per filter, which is needed to obtain the required signal 
to noise, while at the outskirts we allowed shallower exposure.
The increased field of view enables studies that
require high resolution at somewhat larger cluster-centric radii, including weak
lensing measurements, selection of cluster member galaxies for
strong lensing analysis, and galaxy cluster science. 

The WFC3-IR observations were executed within a single orbit,
four images per filter with small box dithers for PSF reconstruction and
to cover artifacts such as the ``IR Blobs'' and ``Death Star'' (WFC3 Data
Handbook; Rajan et al. 2011). We used sampling interval parameter SPARS25.

The subexposures of each filter were reduced and combined 
following the reduction pipeline of our Cycle-20 program GO-13003
(e.g., Sharon et al. 2014).
The WFC3-IR images were treated using a custom algorithm to remove the
``IR Blobs'', and we corrected the ACS images for CTE losses prior to drizzling.
Individual corrected images were combined using the AstroDrizzle package
(Gonzaga et al. 2012) with a pixel scale of $0\farcs03$ pixel$^{-1}$, and drop size of
0.5 for the IR filters and 0.8 for the ACS filters. This approach provides good recovery of the PSF in all
bands and maximizes the sensitivity to detail. All images were aligned
onto the same pixel frame. 
In the final reduced data, the $5\sigma$ limiting magnitudes in the
five filters are 27.4, 27.8, 27.3, 26.8, and 26.5 mag within a
circular aperture of diameter $0\farcs7$, for F435W, F606W, F814W, F110W, and
F160W, respectively. 

A photometric catalog of all the objects in the overlapping ACS and
WFC3 field of view was generated following procedures outlined
in Skelton et al. (2014), and spectral energy distribution (SED) fits
and photometric redshifts derived using EAZY (Brammer et al. 2008). We
note that the fidelity of the photometric redshift is limited by the
small number of filters, nevertheless, the photometric redshifts are
found to be consistent with the available spectroscopic redshifts.

\subsection{Gemini Spectroscopy}\label{s.spec}
The main scientific goal of the \hst\ observations was to facilitate a
detailed lens model of \clustername. Strong lens modeling relies on
constraints from observational evidence of strong lensing, in the form
of multiple images of lensed background sources. The positions and
redshifts are used as local solutions of the lensing equations to
constrain the projected mass density distribution at the core of the
cluster. The accuracy of a lens model strongly depends on the
availability of lensing constraints. The mass
distribution and lensing magnification are sensitive not only to the
accurate identificaitons and positions of multiple images, but also to
the redshifts of these lensed galaxies. This is especially important
when there are few lensed sources identified (Johnson \& Sharon
2016). Lens models that are computed with no spectroscopic  
redshifts as constraints are shown to produce erroneous results (e.g.,
Smith et al. 2009, Johnson \& Sharon 2016); it is therefore critical to
include constraints from at least a few spectroscopically-confirmed
source redshifts.
 
We were awarded 4.5 hours of Band One queue observations with Gemini
Multi-Object Spectrograph \citep[GMOS;][]{GMOS}  
on the Gemini North telescope (GN-2015B-Q-27; PI: Sharon) to secure spectroscopic
redshifts of the secondary arcs that were identified in the new \hst\ data.

Observations of this program
 were executed on UT 2015 Sep 10  and  2015 Nov 6 \& 7. Conditions at the times 
of observation were photometric, with seeing between
$0\farcs5-0\farcs85$. 
The field was imaged by our NOT/ALFOSC monitoring program during the same dark runs,
indicating little variability between these epochs; we
report the $g$-band photometry for reference: 
On 2015 Sep 13.16, g(A)=21.67; g(B)=22.07; g(C)=21.76. On 2015
Nov 07.90, g(A)=21.71 mag; g(B)=21.98 mag; g(C)=21.52 mag.  

For the Gemini/GMOS observations, 
GMOS was configured in macro nod-and-shuffle (N\&S) mode with the R400\_G5305 grating in 
first order and the G515\_G0306 long pass filter. The detector was binned by a factor of 2 in the 
spectral direction and unbinned spatially. Following extensive previous experience using GMOS 
in this mode \citep[e.g.,][]{bayliss2011b,bayliss2014b} we chose a N\&S cycle length of 120~s as a 
balance between achieving good sampling of time variation in the sky and limiting charge trap effects 
by minimizing the number of shuffles in a given integration. 

\begin{figure}
\centering
\includegraphics[scale=0.37]{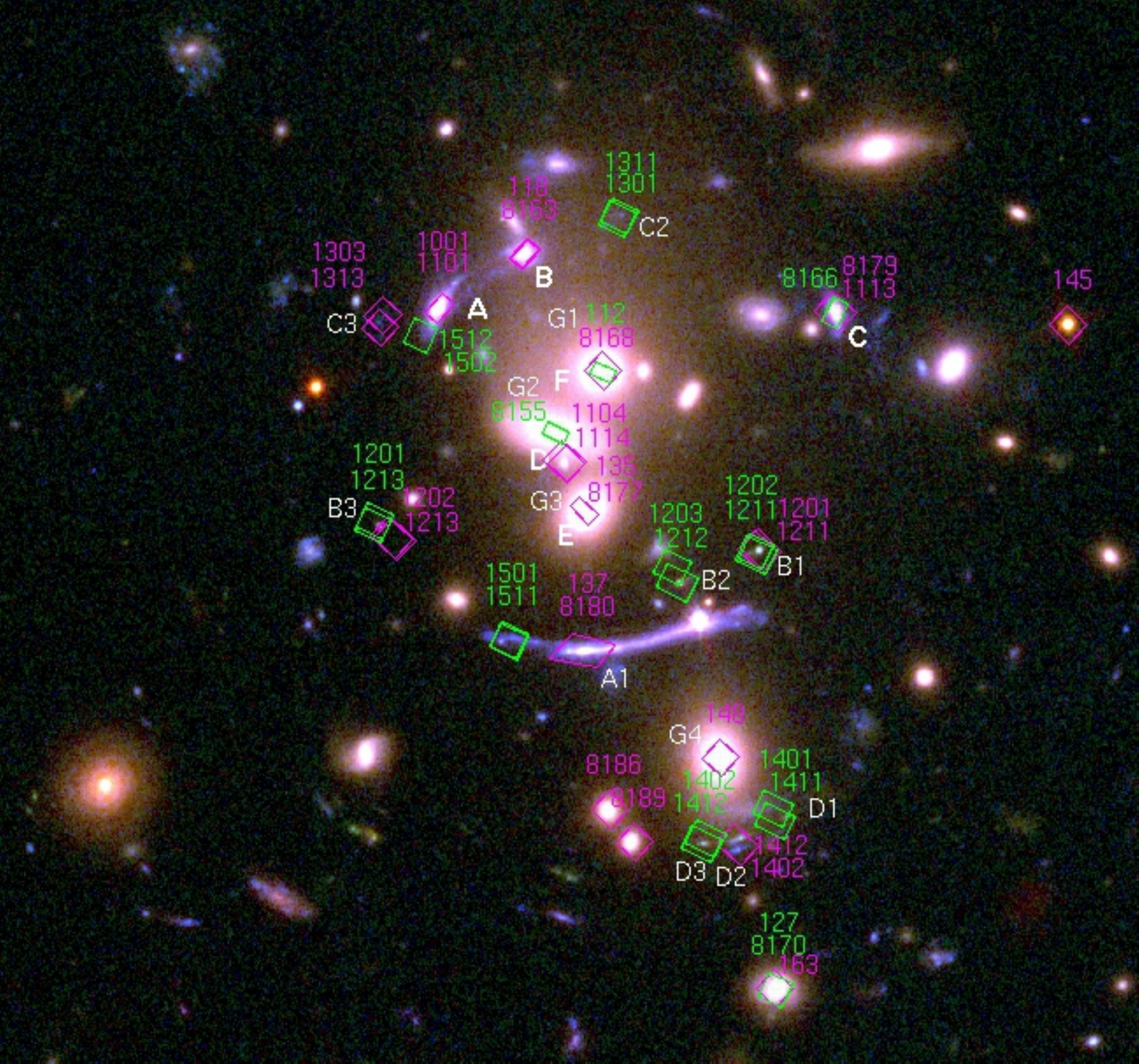}
\caption{The sources that were targeted for Gemini-North Gmos
  multislit spectroscopy are labeled in green (Mask~1) and red (Mask~2). Faint sources were targeted by both masks or in both nod positions, in order to maximize
  their signal to noise.}
\label{fig.masks}
\end{figure}

We designed two multi-object slit masks that preferentially placed slits 
on faint candidate strongly lensed background sources around the core of \clustername\ (see 
Figure~\ref{fig.masks} for mask design and Figure~\ref{fig.hst} for
source IDs). 
Mask~1, at position angle of {65} degrees, targeted lensing candidate images B1, B2, B3, C2, D1, D3, a
faint edge of A1, image C of the
quasar and image A of the host galaxy of the quasar. Mask~2 at
position angle of {47} degrees
targeted arc A1, B1, C3, D2 and quasar images A, B, C, and D. Both
masks targeted cluster member galaxies and other galaxies in the
field.

Slits were placed so as to target high-priority sources at both the original 
pointing position and the offset nod position; this slit strategy is
the same as described in \citet{bayliss2011b}, 
and we refer to that paper for a detailed description. Most slits on each mask were 1\arcsec\ wide, with 
lengths varying from slit to slit. Two slits on each mask had widths
of $0\farcs5$; these were placed on 
the three brightest red galaxies in the core of the cluster to produce higher resolution spectra, which may 
potentially inform stellar velocity dispersion measurements for those galaxies. Each spectroscopic 
mask was exposed twice for 2400~s, with a wavelength dither between the exposures to cover the 
chip gaps in the GMOS detector array. 

We reduced the resulting GMOS spectra using a suite of custom tools that was developed using 
the XIDL\footnote{http://www.ucolick.org/$\sim$xavier/IDL/index.html} package; this pipeline is 
similar to that used in \citet{bayliss2014b}. For N\&S spectra sky subtraction simply requires differencing 
the two shuffled sections of the detector. We first performed this differencing of the raw spectra, and then 
wavelength calibrated, extracted, stacked, and flux normalized spectra from each slit on each of the two 
masks. Flux calibration was performed using an archival standard star. The archival calibration provides 
a reliable relative flux correction, but does not yield an absolute flux calibration. The spectral resolution 
of the final data is R $\simeq 700-1100$ ($270-430$ \kms) for spectra taken through 1\arcsec\ wide slits, 
and R $\simeq 1400-2200$ ($135-215$ \kms) for spectra taken through $0\farcs5$ wide slits.

We summarize the results of the
Gemini spectroscopy observations in Table~\ref{tab.spectroscopy}. Details of the spectroscopic
analysis of the high priority sources are given below. 

\begin{figure*}
\centering
\includegraphics[scale=0.55]{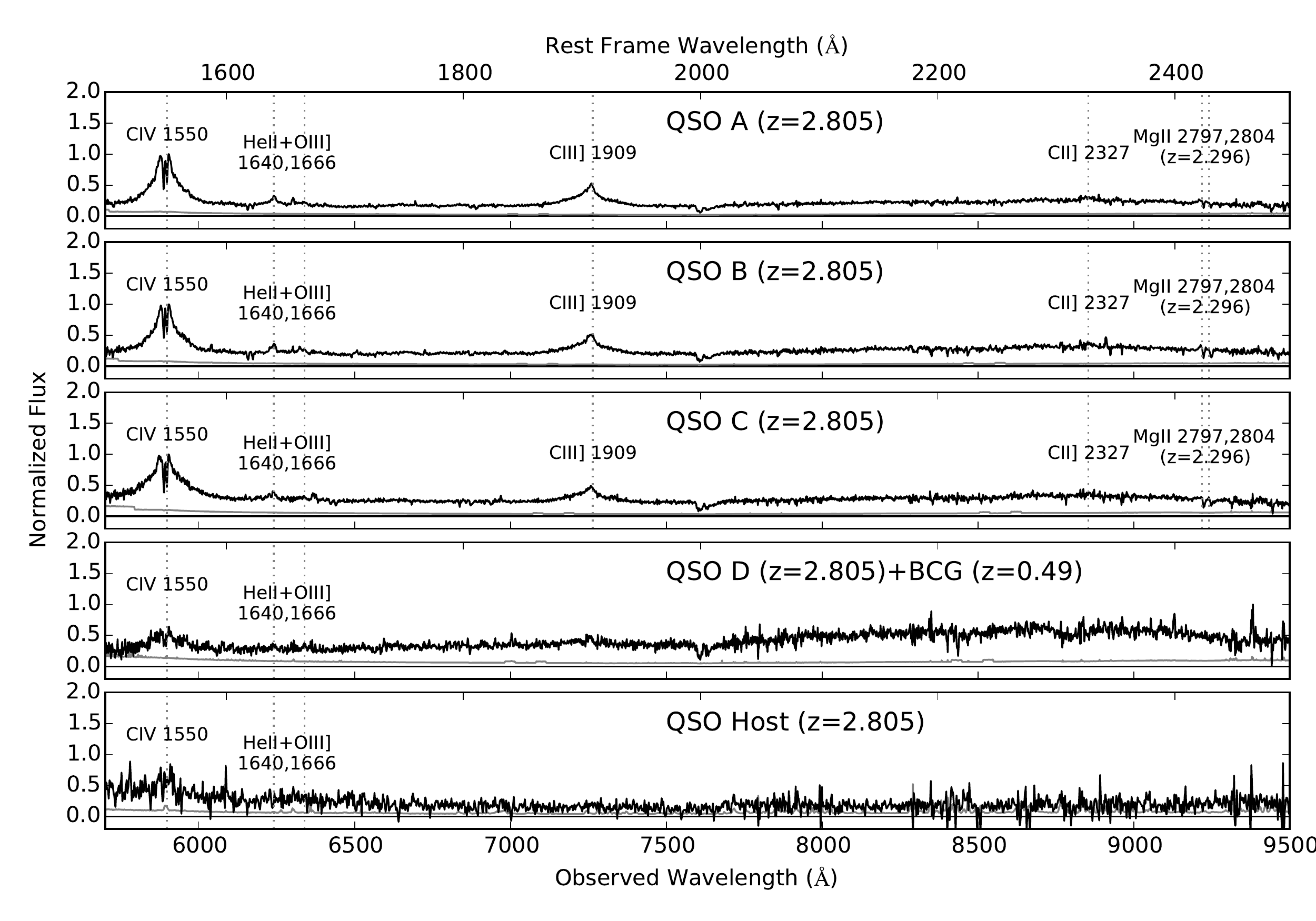}\\
\includegraphics[scale=0.6]{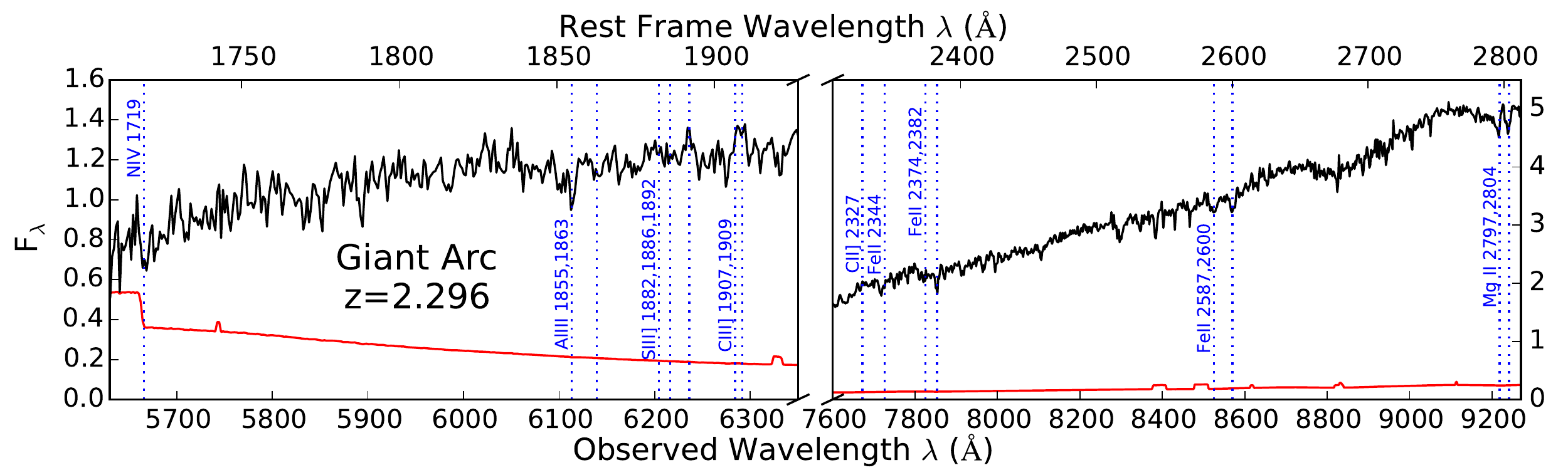}
\includegraphics[scale=0.60]{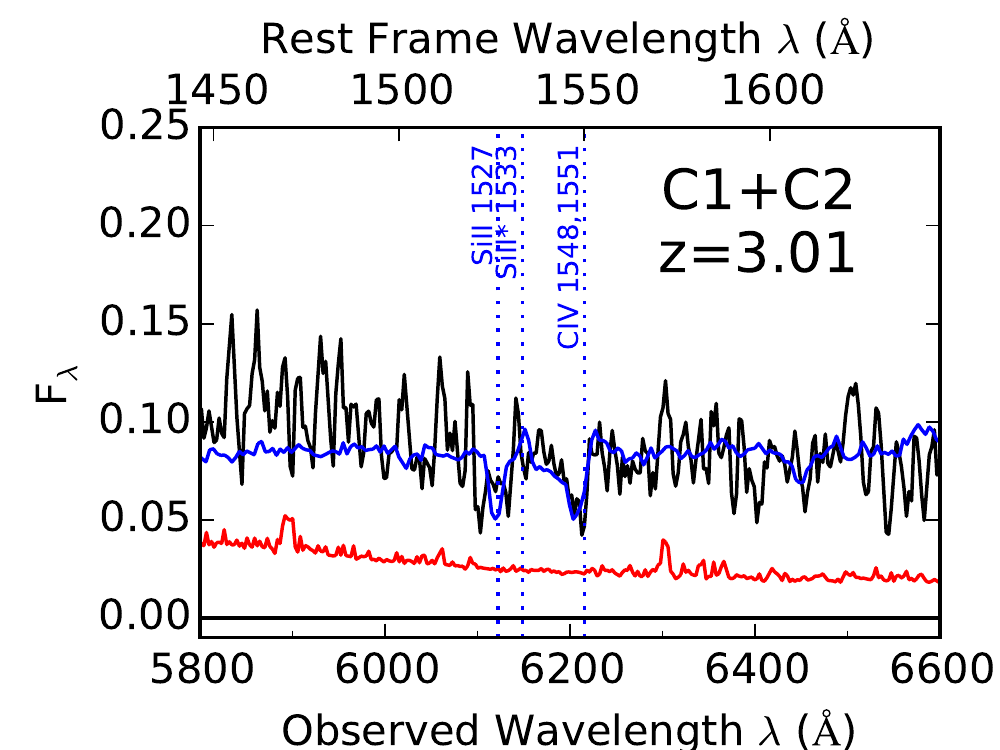}
\includegraphics[scale=0.60]{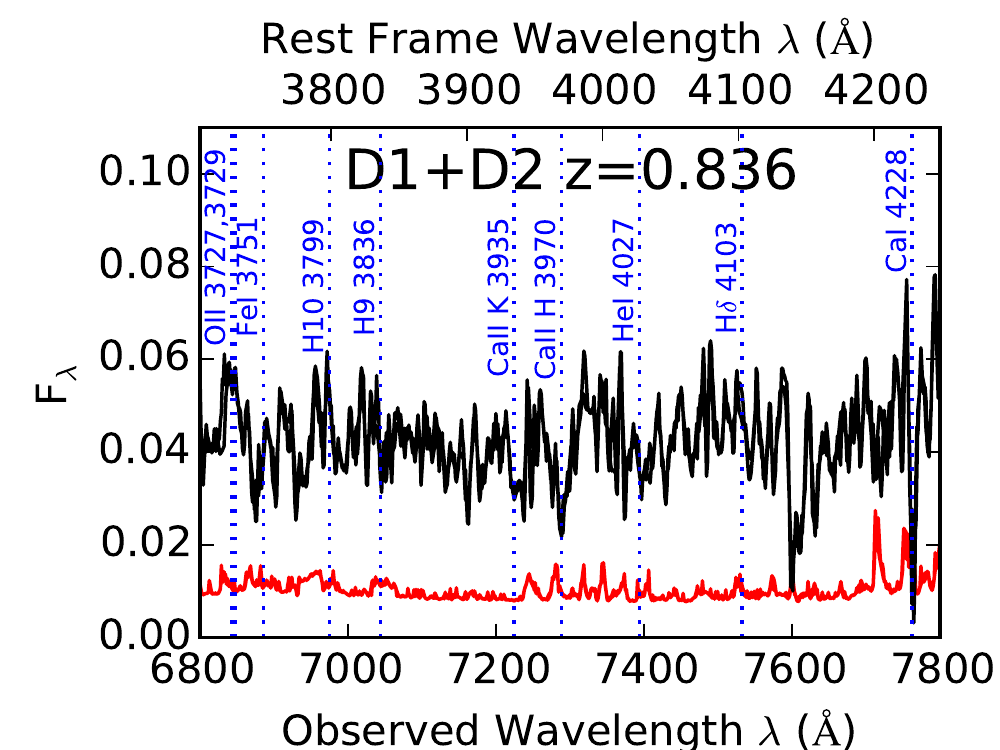}
\includegraphics[scale=0.6]{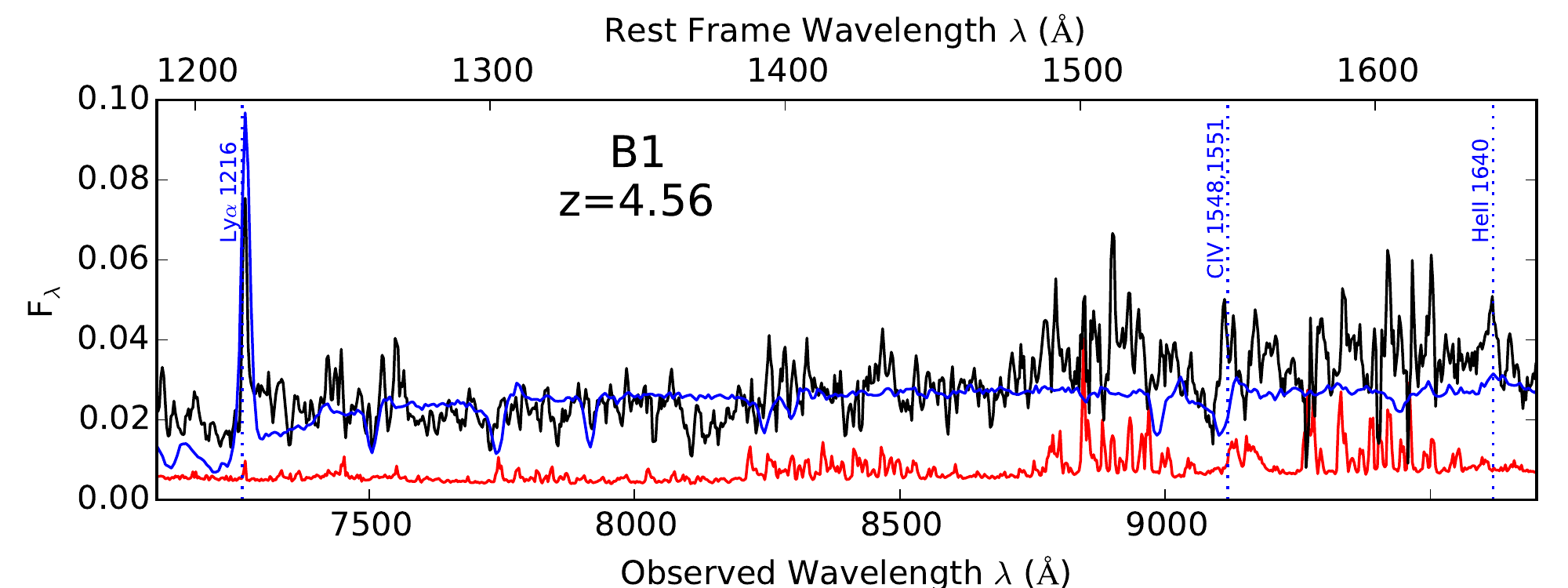}
\caption{Gemini/GMOS spectra of the quasar and arcs A, B, C, D in \clustername. The
  noise level is plotted in magenta. Emission and absorption lines are
  noted. In blue, we overplot the composite stacked spectra of Lyman Break Galaxies
  from Shapley et al. (2003).} 
\label{fig.gmos}
\end{figure*}

\textit{Quasar images}: We refine the redshift measurement of the
quasar, and obtain $z=2.8050\pm0.0006$ from the spectra of images A, B, C of the
quasar (Figure~\ref{fig.gmos}, top panel). We observe emission lines from HeII~1640,
OIII]~1666, [OII]~2470, and CII]~2327, and CIV~1549. We also detect absorption
lines from MgII and other elements at $z=2.296$, from the intervening 
galaxy A1 (see Section~\ref{s.absorber}).  
The spectrum of image D is dominated by light from the foreground cluster
galaxy, however, CIV and CIII emission lines from the quasar can be
detected. A slit targeting the host galaxy of image A of
the quasar (see Figure~\ref{fig.mask} for slit placement) resulted in
low S/N spectrum that is dominated by light from the nucleus.   

\textit{Lensed galaxy A1}: From two slitlets in Mask~2, we confirm the known
arc redshift of $z=$\zarcA\ from ISM absorption lines Fe~II~2344, 2382;
Fe~II~2586,2600; and MgII~2798, 2803. We  
identify weak nebular emission of
Si~III]~1892, and C~III]~1909. The combined spectrum is shown in 
Figure~\ref{fig.gmos}. The slitlet that targeted the faint
region of the arc did not result in sufficient S/N. 

\textit{Lensed galaxy B}: Figure~\ref{fig.gmos} shows stacked spectra
of arc B1 from four slitlets, two in Mask~1 and two in
Mask~2. 
We identify emission lines from 
\Lya, SiII~1260, OI+SiII~1303, HeII~1640, and CIV~1449 at $z=4.56$. The images of
B drop out completely from the ACS/F435W filter, 
which supports this redshift interpretation. Furthermore,  
the photometric redshift analysis obtained for
this source from the five \hst\ bands shows a single high
significance peak around $z_{\rm phot}=4.4$. 
The slits placed on B2 and B3 resulted in too low S/N for an
independent measurement of the redshift. Nonetheless we detect a faint emission
line in these spectra that is consistent with \Lya\ at the same redshift as
image B1.

\textit{Lensed galaxy C}: 
We targeted C2 and C3, with a total on-target exposure time of 2400~s on
each image;
however since these sources are faint, the resulting spectra have low
S/N. A possible absorption line is detected at 6210~\AA.
Interpreting this absorption feature as the CIV~1548,1550 lines, which
are often among the most prominent rest-frame 
UV features in star-forming galaxies, places this source at z=3.01. We
note that this putative spectroscopic redshift is also consistent with
the photometric 
redshift analysis and favored by the lensing analysis. Nevertheless,
given its low certainty we do not consider this a secure spectroscopic
redshift for the purpose of lensing analysis.  

\textit{Lensed galaxy D}: 
We placed slitlets on D1, D2, and D3. As can be seen in
Figure~\ref{fig.masks}, the slits are expected to contain light from
adjacent sources (in particular, blue emission from a nearby
galaxy). The spectroscopic analysis results in a low confidence
redshift of $z=0.837$ for D1 and D2, based on Ca H\&K lines, and no signal in D3. The
photometric redshift probability distribution function is bimodal, with a high significance peak around
$z=4.4$ and low-significance peak around $z=0.74$ for images D1 and
D3. Image D2 shows only one peak at $z=0.74$. Since this is the image
that is most contaminated by blue light from the nearby galaxy, we
argue that this interpretation is consistent with two separate
redshifts for two different background sources. A blue arc at low-z,
consistent with the possible $z=0.837$ that is suggested by the
spectroscopy, and a high redshift source at $z\sim 4.4$ which is likely
the three-imaged lensed source D. Due to the ambiguous redshift
interpretation, we leave this redshift as a free parameter as well,
with upper redshift prior $z<5$ set by the photometric redshift analysis.

\textit{Cluster galaxies}: We measure spectroscopic redshifts of 11 cluster galaxies, including the
central galaxies G1, G2, G3 and G4. These measurements confirm
the published spectroscopic redshifts of G1, G2, G3 from the NOT (Dahle et al. 2013). 
In the spectrum of galaxy G3 we detect weak C~IV emission
at $z$=\zsource\ from the embedded quasar image E. In
Table~\ref{tab.spectroscopy} we also list the redshifts of three
galaxies with SDSS-DR9 spectroscopy that are within projected radius
of 1500 h$^{-1}$ kpc from the BCG. 
From these 14 members we measure a cluster redshift of
$z=0.4897\pm0.0032$, and a 
 velocity dispersion of
\vdisp, using the Gapper estimator
(Beers et al. 1990). The uncertainties on the velocity dispersion are
calculated as $\pm0.91\sigma_v/\sqrt{n-1}$, where $n$ is the number of
galaxies, following Ruel et al. (2014). Using the mass scaling relation in Evrard et
al. (2008), the velocity dispersion translates to a dynamical mass of \Mdyn.

\textit{Other galaxies}: Table~\ref{tab.spectroscopy} also lists the coordinates and
the spectroscopic
redshifts of other background (i.e., behind the cluster) and
foreground galaxies in the field that were measured from these data.

\input{spectable.tex}

\subsection{SWIFT X-ray Observations}\label{s.swift}
\clustername\ was observed at X-ray wavelengths by the \Swift\ X-ray
telescope (XRT) as part of a monitoring program using University of
Michigan time (PI: Sharon). 
Observations were taken approximately every
six weeks over seven epochs between 2015 September 16 and 2016 June
29, with a combined exposure time of
90.5~ks. The typical exposure time was 15 ks per epoch with the
exception of epoch~1 that was observed for $\sim$10 ks  (see Table~\ref{swift:obs}). 
The hard X-ray radiation ($< 3$~keV) varies during this time by up to a
factor of three between epochs with the lowest and the highest counts
per second, confirming the variable nature of 
the quasar at these wavelengths. However,  
the \Swift\ XRT resolution (see below) is not
sufficient to robustly resolve the three brightest images of the lensed quasar. A
decomposition analysis of the variable emission is beyond the scope of this
paper, and will be presented in future work. 
 Here, we present the co-added data from the first seven epochs, and
 analyze the X-ray emission from the cluster hot gas.   

The data were reprocessed using the HEASOFT v.~6.17 and the most
up-to-date version of CALDB, accessible via remote server.  New
Level~2 event files were created using the tool \texttt{xrtpipeline}.  
We used the XRT Data Product
Generator\footnote{http://www.swift.ac.uk/user\_objects/index.php} to
combine images from different epochs, and for astrometric
measurements. We verified the absolute astrometric solution by matching the
coordinates of a bright X-ray star at  [RA,
Dec]=[335.39729, 27.707253],
with its optical counterpart from the SDSS.

\begin{figure}
\centering
\includegraphics[scale=0.55]{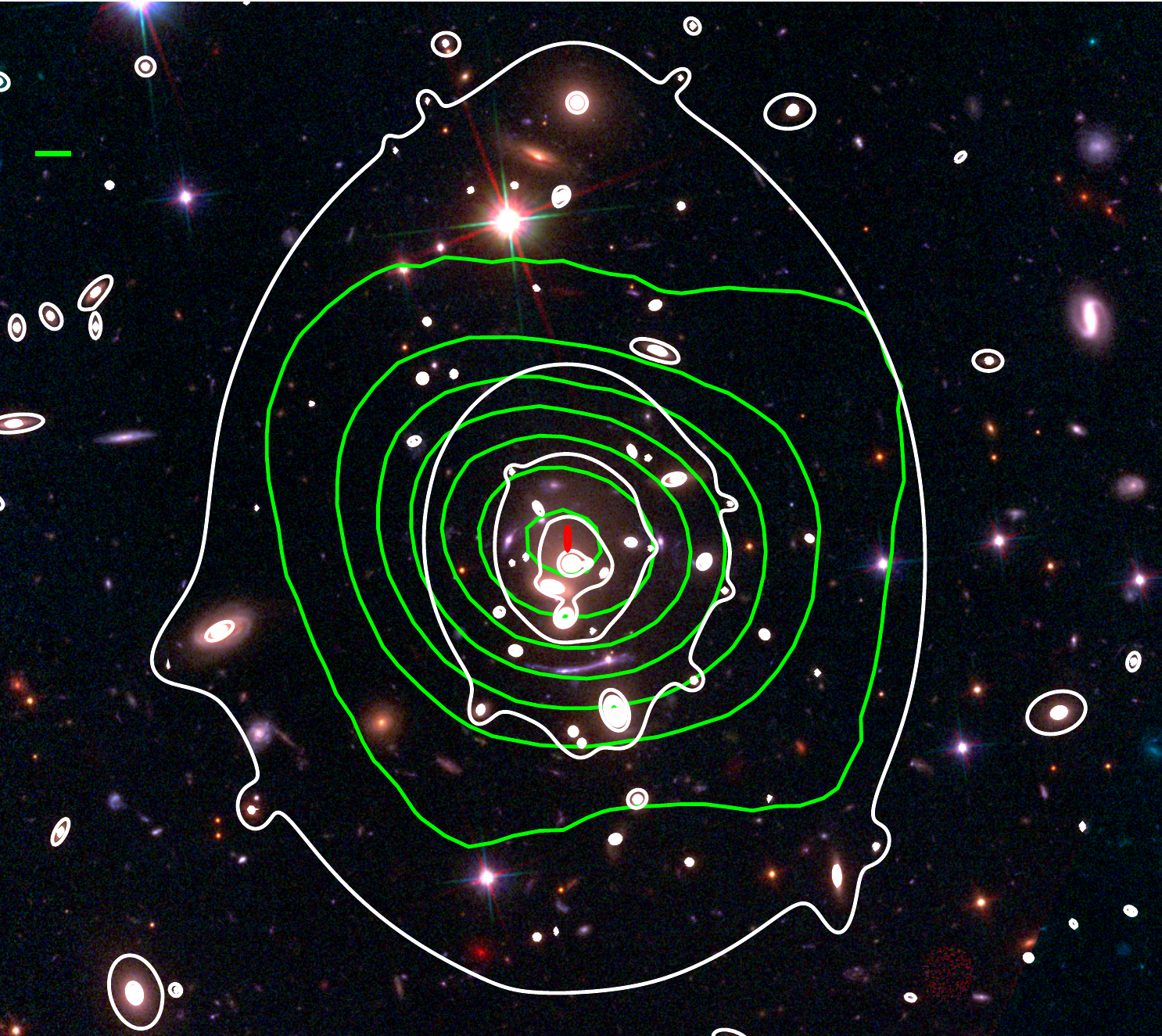}
\caption{A three-color HST image of SDSSJ2222 with \Swift\
  X-ray contours from the energy range $0.1-2.4$~keV overlaid in
  green, and the lensing mass contours in white. The red ellipse
  indicates the statistical uncertainty on the centroid of the
  main cluster halo component from the lensing analysis (see
  Section~\ref{s.lensing} and Table~\ref{table.lensmodel}).  
The X-ray emission was smoothed with a Gaussian kernel
  matching the \Swift\ PSF. Contours are linearly spaced.  The thick, horizontal green bar
  in the upper left-hand corner of the image is $3\farcs5$ in length,
  which is size of the astrometric uncertainty.}   
\label{fig.xray}
\end{figure}

The \Swift\ XRT has a PSF of $18''$ at 1.5~keV; 
at this resolution, the X-ray radiation from the
images of the lensed quasar is blended with the diffuse emission from the 
hot cluster X-ray gas. 
Nevertheless, their contribution is
wavelength dependent, with the soft X-ray
radiation ($\lesssim 3$~keV) dominated by the cluster
emission, and the hard X-ray photons attributable to the quasar.  
In order to separate the cluster emission from the quasar emission, we
 co-add the emission in the $0.3-3.0$~keV range from all seven
 epochs. The resulting X-ray contours are over-plotted on the optical
 image in Figure~\ref{fig.xray}. 
We find that the soft X-ray emission is centered on [RA,
Dec]=[335.53613, 27.760667],
with a 90\%
confidence error radius of $3\farcs0$ (see Evans et al. 2014 for more information on the way the astrometric
position is determined).  This position is in excellent agreement
with
the center of the main cluster halo component as derived from the
lensing analysis (Section~\ref{s.lensing}). The two centroids
are $1\farcs11$ in projection, a distance which is within the
$3\farcs$ XRT astrometric uncertainty.
Nevertheless, since the XRT Data Product
Generator is not optimized for measurements of extended sources the
uncertainty may be underestimated.  

For an estimate of the cluster mass, we measure the
background-subtracted X-ray flux, from the  co-added data of the 
seven epochs, in the energy range 0.1-2.4 keV, within an aperture of
radius 1~Mpc centered on the cluster; we assume that this radius
corresponds roughly to R$_{500}$ (e.g., Mantz et al. 2010), and that in this energy range the
X-ray radiation is dominated by the cluster with negligible
contamination from the background quasar. 
To account for errors due to the unknown
 gas temperature we consider  
 a range of gas temperatures between kT=[$3.06 - 9.67$] keV, and find
 luminosities in the range $L{\rm x}=[1.0-1.4] \times10^{44}$~ergs
 s$^{-1}$. The luminosity is within the range expected for clusters
 with similar velocity dispersion (Xue \& Wu 2000).
 We use  the Mantz et al. (2010) $M-L_{\rm X}$ relation to
 estimate the cluster mass, \Mxraylowhi. The X-ray-inferred mass estimate is in line
with the lensing mass measurement and with the dynamical mass.  
A more
robust measurement of the X-ray mass will be enabled with higher
resolution \textit{Chandra} data, with which the emission from the cluster gas
and the background quasar can be spatially disentangled.
\\
\\

\begin{deluxetable}{lccc}
\tablecaption{\textit{Swift} Observation Information}
\tablecolumns{4}
\tablewidth{0pt}
\tablehead{
\colhead{OBSID} &
\colhead{Start Date} &
\colhead{Exp. Time (ks)} &
\colhead{Epoch}
}
\startdata
00034046001 & 2015-09-16 & 2.4 & 1 \\
00034046002 & 2015-09-27 & 6.2 & 1\\
00034046003 & 2015-11-06 & 12.8 & 2 \\
00034046004 & 2015-12-18 & 9.0 & 3 \\
00034046005 & 2015-12-20 & 3.8 & 3 \\
00034046006 & 2016-01-30 & 13.7 & 4\\
00034046007 & 2016-04-13 & 15.6 & 5\\
00034046008 & 2016-05-16 & 14.1 & 6\\
00034046009 & 2016-06-28 & 3.3 & 7\\
00034046010 & 2016-06-29 & 9.5 & 7\\
\enddata
\label{swift:obs}
\end{deluxetable}

\section{Strong Lensing Analysis}\label{s.lensing}
\subsection{Multiple images and lensing constraints}\label{s.arcs}
The lens model of \clustername\ relies on observational strong lensing
evidence, in the form of multiply-imaged galaxies. The multiband
\hst\ images are uniquely useful for the task, owing to their high
resolution and broad wavelength coverage that allow identifying
multiple images of individual background sources by their color and
morphology. 
In Dahle et al. (2013) we identified six images of one
background quasar in imaging data from the Nordic Optical
Telescope. We confirmed five of these images and secured their redshift
 through spectroscopy. The sixth image was predicted by the
preliminary lens model and identified in the data after modeling and
subtracting the light of the cluster galaxies at the core of the
cluster; Dahle et al. (2013) provide strong evidence for the presence
of the sixth image. 
 The \hst\ images confirm the sixth image as a counter image
of the quasar, with a point-like PSF and similar colors to the other
quasar images. These images are labeled A, B, C, D, E, F in Figure~\ref{fig.hst}.
A second lensed galaxy A1, at $z=\zarcA$ (Dahle et al. 2013, Stark et
al. 2013), is distorted by the cluster and appears
as a blue giant arc south of the cluster center. 
The new \hst\ data reveal substructure in the giant arc A1, but do not
lead to an identification of a counter image of this galaxy. We 
interpret this giant arc as a likely result of source-plane caustics
that bisect the galaxy or pass very close to it, resulting in  high magnification in the tangential
direction (see Section~\ref{s.source}).

We identify three secure strongly-lensed galaxies with multiple images in the new \hst\
data.  

\textit{Source B} has three multiple images with unique
color, morphological resemblance, and the expected
parity. We measure a spectroscopic redshift of $z_{\rm B,spec}=\zarcB$ using 
GMOS on Gemini North (see Section~\ref{s.spec}). 

\textit{Source C} is a faint source observed as three images with similar
lensing configuration as the three brighter quasar images north of the
cluster core. Due to the low surface brightness of C1, C2, and C3, we
were not able to obtain a secure spectroscopic redshift. The
photometric redshift, spectroscopy, and lensing geometry are all
consistent with it being at
$z_{\rm C,phot}\sim3$  (see Section~\ref{s.spec}). We leave the redshift of this source as free parameter in
the lensing analysis, with broad priors based on the photometric
redshift analysis, $2.0\le z_C\le4.0$. 
We expect
that further counter images of this source would be too faint and
embedded in the light of the bright cluster galaxies to be detected in
the existing data.
 
Three images of \textit{source D} appear in the WFC3/IR bands, south of a cluster-member galaxy in the
south part of the cluster core.
As described in Section~\ref{s.spec}, we were unable to measure a
secure spectroscopic redshift for this source.  We leave the redshift of this source as free parameter in
the lensing analysis, with broad priors based on the photometric
redshift analysis, $3.8\le z_D\le 5.0$. 

We identify other candidates of lensed galaxies, however, these are
not robustly confirmed as strong lensing features and thus are not
used as constraints in the lens model. 

We use the positions of the six quasar images, arcs A1, B1-3, C1-3 and
D1-3 to constrain the lens model. Resolved emission
knots and substructure in the host galaxy of the quasar and B1-3
 are also used as additional positional constraints. 
The redshifts of the
quasar and sources A and B are used with no uncertainty, while the redshifts
of source C and source D are left as free parameters with broad priors
set by the probability distribution functions of their
photometric redshifts.

\subsection{Strong Lens Model}\label{lensmodel}
The lens model is computed using the public software \Lenstool\ (Jullo
et al. 2007). Lenstool relies on a `parametric' modeling algorithm, in
which the mass distribution is assumed to be a combination of a number of
halos, each described by a set of parameters. The software 
uses Markov Chain Monte Carlo (MCMC) procedure to sample the parameter
space, determine the best set of parameters that minimize the scatter
between the observed and predicted positions of multiply-imaged lensed
galaxies, and determine their uncertainties. 

\clustername\ is modeled with one cluster-scale halo, plus
galaxy-scale halos. Each of these halos is
modeled as a Pseudo-Isothermal Elliptical Mass Distribution (PIEMD;
also known as dual Pseudo Isothermal Elliptical Mass Distribution,
El{\'{\i}}asd{\'o}ttir et al. 2007). The parameters of this mass
distribution are positions $x$ and $y$; ellipticity,
$e=(a^2-b^2)/(a^2+b^2)$, where $a$ and $b$ are the semi-major and
semi-minor axes, respectively; position angle $\theta$, measured
north of west; core radius $r_{\rm core}$; cut radius $r_{\rm cut}$; and effective
velocity dispersion $\sigma_0$. We allow all the parameters of the
cluster-scale halos to vary, except for $r_{\rm cut}$, which, for a typical cluster, 
is much larger than the radius in which lensing evidence can be
found and thus cannot be constrained by the model. We fix the
cluster-halo $r_{\rm cut}$ at 1500 kpc.  

Cluster-member galaxies are selected from a color-magnitude diagram,
as those with colors that place them on the cluster red sequence
(Gladders \& Yee 2000). We
note that some galaxies at this redshift may not be quiescent and
therefore fall off of this relation. However they are not a dominant
component at the core of the cluster (e.g., Fairley et al. 2002). 
The cluster galaxies are also modeled as PIEMDs, with morphological
parameters ($x$, $y$, PA, $\theta$) fixed to their observed values as
measured from the \hst\ data in ACS/F814W. $r_{\rm core}$, $r_{\rm cut}$ and
$\sigma_0$ are assumed to correlate with the luminosity of each galaxy (see Limousin
et al. 2005 for a description of the scaling relations).

The slope parameters of five galaxies at the center of the
cluster are allowed to deviate from the scaling relation.  The
lensing potential of the  three brightest galaxies near the core of
the cluster is responsible for the appearance of the three fainter
images of the quasar -- D, E, and F. In a close inspection of
the galaxies near images D and F, we find that the peak of
surface brightness is not aligned with the center of the light
distribution of these galaxies, implying a more complex projected mass
distribution than that of a single elliptical halo, at least of its
stellar mass component. This may be due to
the merger history of these galaxies (e.g., Lidman et al. 2013; Lavoie
et al. 2016) or a
projection effect. 
We therefore model each of these galaxies as a combination of two
halos. One halo has its $x$, $y$ parameters fixed to the
center of the extended light distribution of the galaxy, its
ellipticity and position angle follow those of the light distribution,
and the other
parameters allowed to vary. The second halo is centered on the peak
surface brightness, with circular 
symmetry, vanishing core radius, and $\sigma_0$ and $r_{cut}$ set as
free parameters. 

The distribution of the intracluster light is observed to by more
extended in the 
North-South direction (Figure~\ref{fig.hst}), which would be consistent with a young
dynamical age for the cluster. However, the deep combined NOT and
Gemini images (Dahle et al. 2015) indicate  
considerable Galactic cirrus in the field, which is difficult to disentangle from
intracluster light at the very faintest surface brightness levels.   

Although we find that some of the free parameters are not sensitive to
the positional lensing constraints, we allow these parameters to vary in order to
encompass the full range of statistical uncertainties, and investigate
their affect on the time delay of the quasar images.

\begin{figure}
\centering
\includegraphics[scale=0.36]{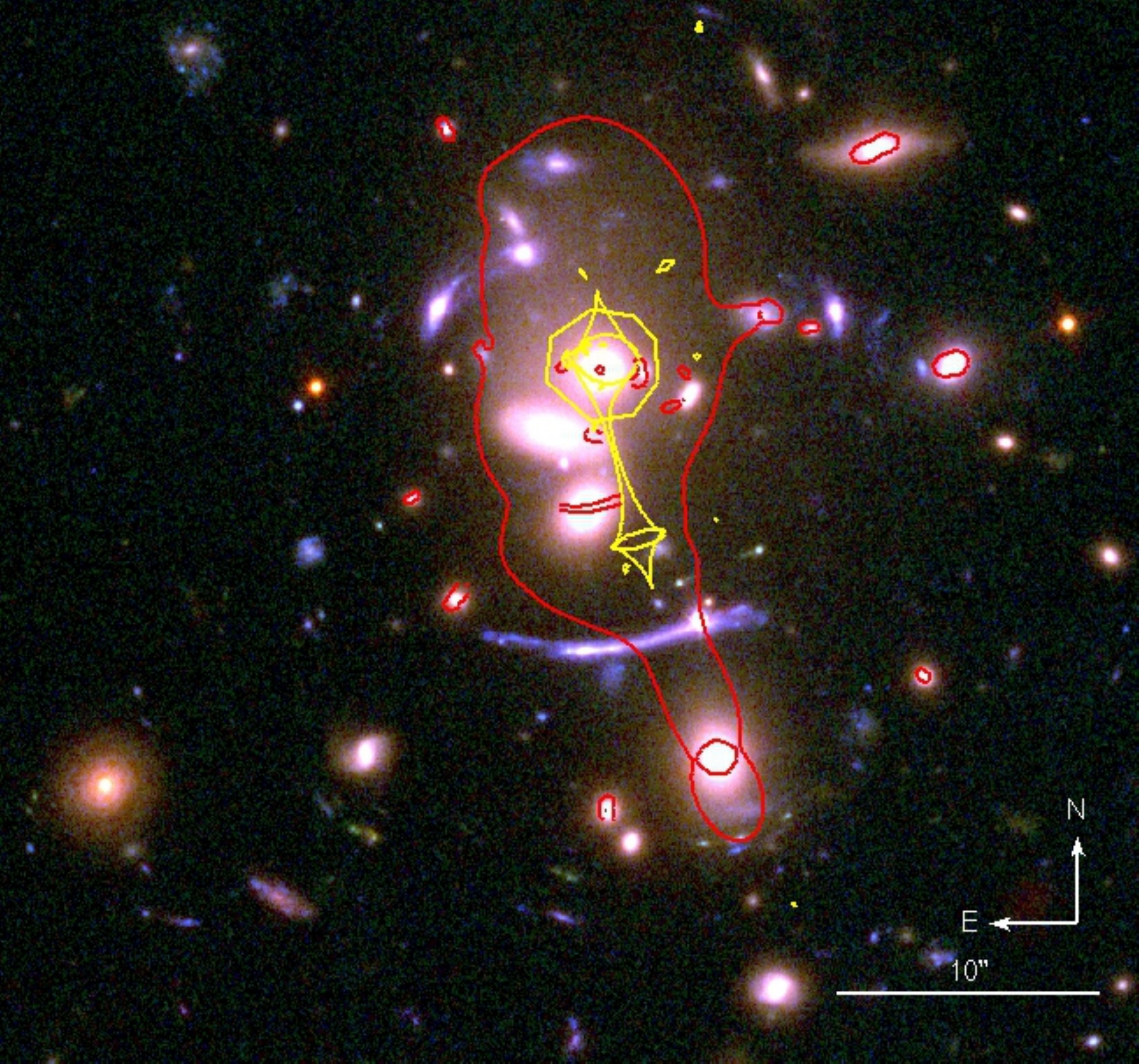}
\caption{The critical curves for a source at $z=\zQSO$ from our
  best-fit lens model are over-plotted on a color-composite \hst\
  image of \clustername. 
}
\label{fig.lensmodel}
\end{figure}

In Table~\ref{table.lensmodel}, we list the lens model parameters and
their uncertainties, including the time delay constraints (95\%
confidence limit from Dahle et al. 2015). We plot the critical curves
from the best-fit model in Figure~\ref{fig.lensmodel}, for a source at $z=$\zsource.
The best-fit model has an image-plane RMS
of \implanerms. We note that
since the RMS was computed from the predicted positions of the  same
images that were used as 
constraints, it is not an unbiased indicator of the model fidelity
(Johnson \& Sharon 2016). 

\begin{figure*}
\centering
\includegraphics[scale=0.29]{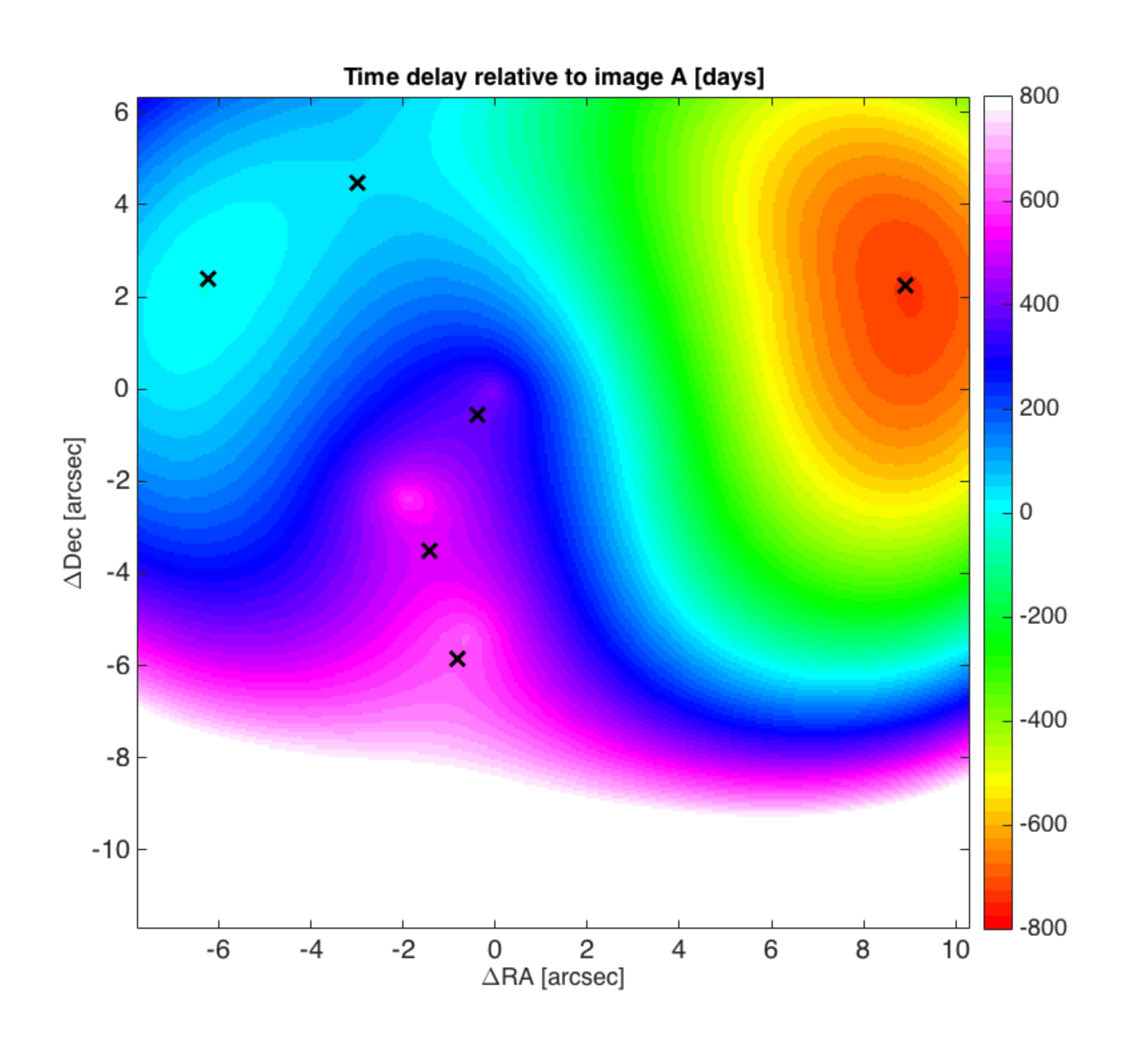}
\includegraphics[scale=0.29]{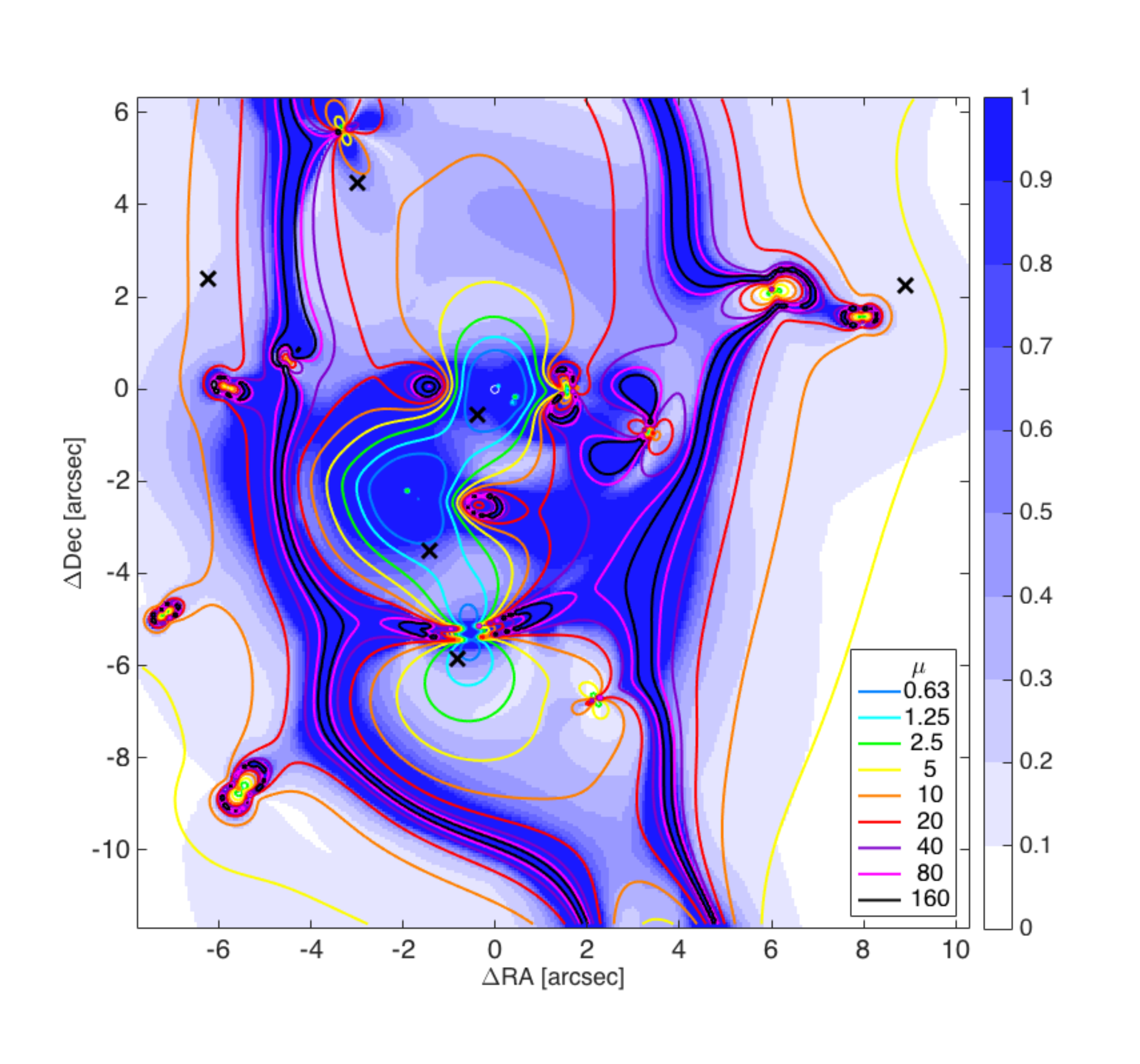}
\caption{ (a) \textit{Left}: Excess arrival time surface (Fermat potential) for
    light emitted from the source position in the source plane at $z=$\zsource, and
    traverses the lens plane at $z=$\zcluster. The best-fit model,
    from which we compute this time surface,
    takes into account the 95\%  confidence interval of the measred time delays
    from \dahleB.
 The excess arrival time is computed
    relative to image A of the quasar, and given in days. 
(b)~\textit{Right}: The magnification map of a source at $z=$\zsource\ from the best-fit
  model is shown in contours. The colormap background gives the
  relative uncertainty in each point, $\Delta\mu/\mu$.
In both panels, the
   positions of the six images of the quasar are marked, and
   coordinates are indicated in arcsec relative to  [RA,
    Dec]=[335.535745, 27.7598861].
\label{fig.magnific}
}
\label{fig.fermat}
\end{figure*}

The time delay between the images of the quasar can be measured from the
arrival time surface (e.g., Schneider 1985), 
\begin{equation}\label{eq.dt}
\tau(\vec\theta,\vec\beta) = \frac{1+z_l}{c}\frac{D_{l}D_{s}}{D_{ls}}\bigg[\frac{1}{2}(\vec\theta-\vec\beta)^2-\psi(\vec\theta)\bigg],
\end{equation}
where $\vec\beta$ is the source location, $\vec\theta$ is a coordinate
in the image plane, $z_l$ is the lens redshift, $D_{l}$ and $D_{s}$
are the distances from the observer to the lens and to the source,
respectively, $D_{ls}$ is the distance from the lens to the source,
and $\psi$ is the lensing potential.  Figure~\ref{fig.fermat}a shows the Fermat potential of the
best-fit model, with the positions of the observed quasar images
overplotted. Multiple images occur in
stationary points in this potential, i.e., maxima, minima, and saddle
points. The lens model successfully predicts the formation of all the observed
quasar image as well as three additional demagnified
images, each within $0\farcs1$ of the  center of galaxies G1, G2, and G3, at
the extrema points of the Fermat potential. 
However, these images are predicted to be several magnitudes
fainter than the faintest observed image of the quasar, with
$29-34$ mag in the F435W band, and thus we do
not expect to be able to detect them in the current data.

We report the predicted arrival time in days relative to image A of
the quasar, $\Delta t=\tau(\vec\theta,\vec\beta)-\tau(\vec\theta_{\textrm{A}},
\vec\beta)$. 
As can be seen in Equation~\ref{eq.dt}, the Fermat potential depends on the
source position, $\vec\beta$. It is in fact very sensitive to small
changes in the exact value of $\vec\beta$. We therefore follow the
procedure described in Sharon \& Johnson (2015), and take the source
plane scatter into account when computing the uncertainties of the
time delay of the quasar images.  The best-fit time delays and their
uncertainties are listed in Table~\ref{tab.magnific}. 

Time delays are not implemented as constraints at this point. We
derive a lens model with no prior on the time delays, and later
confront the model with the measured time delays from Dahle et
al. (2015) in a posterior analysis -- see Section~\ref{s.posterior}. 
After applying the observational time delay constraints on the
posterior distribution, we find that parts of the parameter space are
excluded. 

\begin{deluxetable*}{clccccccc} 
 \tablecolumns{9} 
\tablecaption{Best-fit lens model parameters \label{table.lensmodel}} 
\tablehead{\colhead{No. }   & 
            \colhead{Component }   & 
            \colhead{$\Delta$ RA ($\arcsec$)}     & 
            \colhead{$\Delta$ Dec ($\arcsec$)}    & 
            \colhead{$e$}    & 
            \colhead{$\theta$ (deg)}       & 
            \colhead{$r_{\rm core} $ (kpc)} &  
            \colhead{$r_{\rm cut}$ (kpc)}  &  
            \colhead{$\sigma_0$ (km s$^{-1}$)}             } 
\startdata 
1&Cluster halo                    & \PxA       & \PyA       & \PeA  & \PthetaA        &\PrcA     & \PcutA    & \PsigmaA  \\ 
2&G1 halo          & \PxB       & \PyB       & \PeB  & \PthetaB        &\PrcB     & \PcutB    & \PsigmaB  \\ 
3&G1 core          & \PxC       & \PyC       & \PeC  & \PthetaC        &\PrcC     & \PcutC    & \PsigmaC  \\ 
4&G2 halo          & \PxD       & \PyD       & \PeD  & \PthetaD        &\PrcD     & \PcutD    & \PsigmaD  \\ 
5&G2 core          & \PxE       & \PyE       & \PeE  & \PthetaE        &\PrcE     &  \PcutE    & \PsigmaE  \\ 
6&G3 halo          & \PxF       & \PyF       & \PeF  & \PthetaF        &\PrcF     &  \PcutF    & \PsigmaF  \\ 
7&G3 core          & \PxG       & \PyG       & \PeG  & \PthetaG        &\PrcG     &  \PcutG    & \PsigmaG  \\ 
8&G4                  & \PxH       & \PyH       & \PeH  & \PthetaH        &\PrcH     &  \PcutH    & \PsigmaH  \\ 
9&G5                  & \PxI       & \PyI       & \PeI  & \PthetaI        &\PrcI     &  \PcutI    & \PsigmaI  \\ 
 &L* galaxy  & \nodata & \nodata & \nodata & \nodata &  [0.15]  &     [50]&  [130]  \\

\enddata 
 \tablecomments{The coordinates are given in arcseconds measured East and
   North of the core of galaxy G1, at [RA, Dec]=[335.535745,
   27.7598861].  All the mass components are parameterized as PIEMD,
   with ellipticity expressed as $e=(a^2-b^2)/(a^2+b^2)$. $\theta$ is
   measured North of West. Error bars are inferred from the MCMC
   optimization and correspond to 1$\sigma$. 
   Parameters that were not optimized are listed in
   square brackets. The
   location and the ellipticity of the matter clumps associated with
   cluster galaxies were kept fixed according to their
   light distribution, and the other parameters determined through
   scaling relations (see text).}
\end{deluxetable*}

\section{Results and Discussion}\label{s.results}
\subsection{Time Delays}\label{s.posterior}
Observational measurement of the time delays between the images of the
quasar can provide valuable constraints on the lens model, as the
arrival time is sensitive to the lensing potential. 
\clustername\ gives us a unique opportunity to obtain observational
constraints on the time delay from six images of the same background
quasar, three of which appear close to the cluster core, in close
proximity to cluster galaxies. In Dahle et al. (2015) we report
on the measurement of time delays between the three bright images of
the quasar, \tab=\timeABobs\ days, and \tac=\timeACobs\ days (all the
time delays are measured as excess arrival time relative to image A of
the quasar). 
Our basic lensing analysis does not use the time delays as
constraints, and is done strictly without any a priori knowledge of
the time delays. 

We now confront the lens model with the time delay observations. The
basic lensing analysis predicts that the arrival time is shortest for image
C of the quasar, followed by images A,B,F,D,E. Quantitatively, we find
\tab=\timeAB\ days, and \tac=\timeAC\ days, in
good agreement with the observed measurements of \dahleB. 

\begin{figure*}
\centering
\includegraphics[scale=0.4]{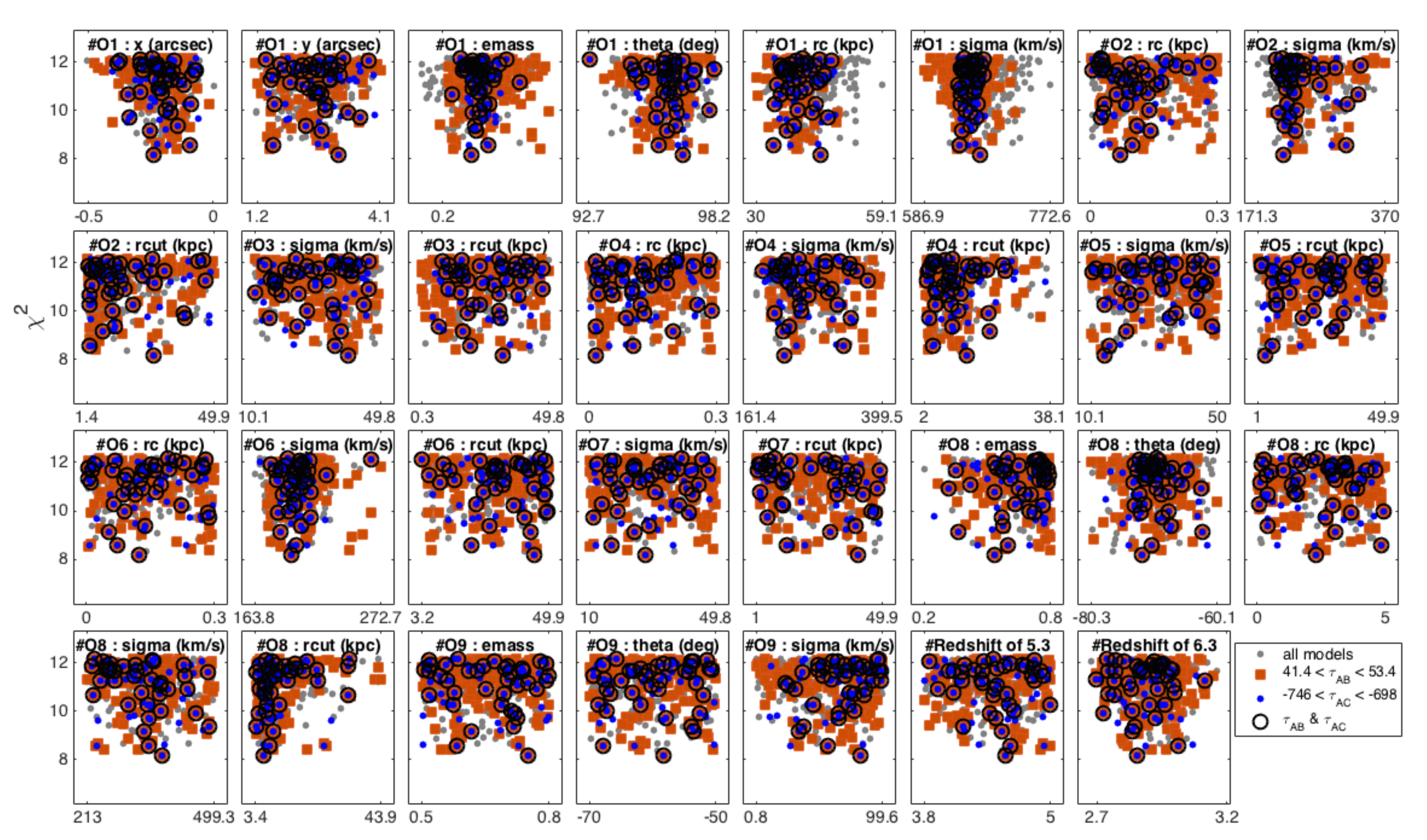}
\caption{ The goodness of fit plotted against the parameter value, for
  sets of parameters from the MCMC analysis. The goodness of fit
  is estimated via the $\chi^2$ value, computed by \lenstool\
 as the scatter between observed and predicted image positions. Models
 that predict  \tab\ in the range \timeABobs\ days are plotted in
 red squares; models that predict \tac\ in the range
 \timeACobs\ days are plotted in green circles; models that satisfy both
 criteria are circled in black.  
All other models are plotted in gray
 circles.
 The observed time delay \tac\ has constraining power over
 the parameters of the main cluster
 halo, mainly the normalization $\sigma_0$, which is correlated with
 the overall mass of the cluster, and the ellipticity $e$. 
}
\label{fig.parameters}
\end{figure*}

Next, we use the observed time constraints and their 95\% confidence
limits to further constrain the parameter space. We select the sets of
parameters from the MCMC sampling that result in lens models with $\chi^2$
in the range [$\chi^2$,$\chi^2+4.5$]. We consider these models as
producing reasonable scatter in the predicted vs. observed positions
of images of the lensed galaxies, and their parameters are drawn from
a range larger than the 1$\sigma$ confidence interval of the
parameter space of well-constrained parameters, as sampled by the MCMC
process.  
Models with larger $\chi^2$ were rejected.
We then compute the Fermat potential for each one of these
sets of parameters (Equation~\ref{eq.dt}), assuming that the quasar
source position $\vec\beta$ is at the
mean of the predicted source positions of the six quasar
images. We compute the excess arrival time relative to image A of the
quasar, i.e.,
the predicted time delay for each of the images. 
We identify the models that predict time delays \tab\  and
\tac\ within the 95\% confidence limit of the observed 
values of \dahleB.  In Figure~\ref{fig.parameters}, we plot the positional $\chi^2$
against each of the parameters of the main cluster halo, and
color-code the models that predict either \tab\ in the range
\timeABobs\ days,
\tac\ in the range \timeACobs\ days, or both.
As can be seen in Figure~\ref{fig.parameters}, while models that predict
the observed \tab\ span the entire parameter space, the
measured time delay between image C and A, \tac,
has good constraining power over some of the parameters, mainly the
overall mass of the main cluster halo ($\sigma_0$), and its ellipticity
($e$). 

\begin{figure*}
\centering
\includegraphics[scale=0.4]{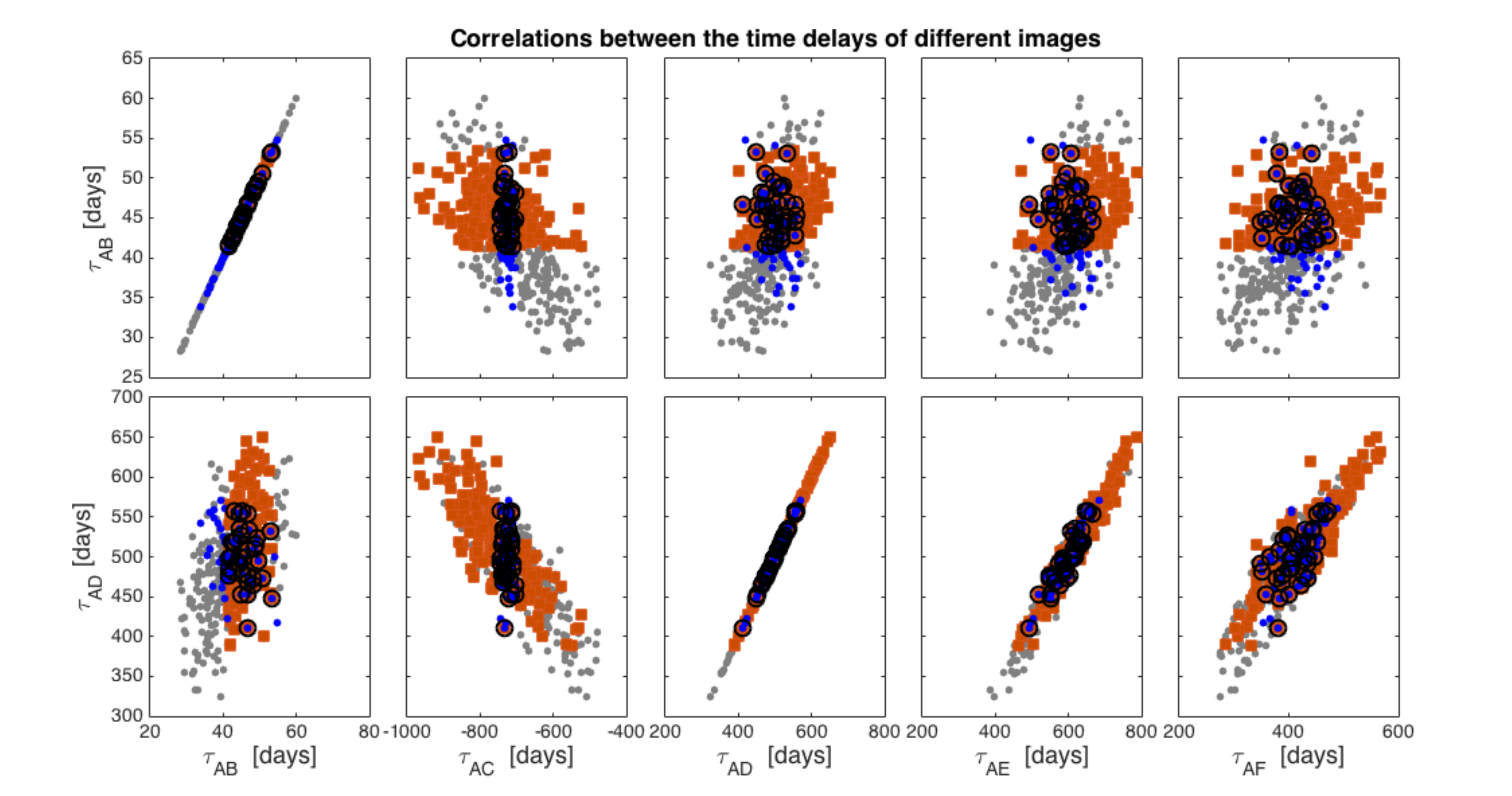}
\caption{Correlations between the predicted relative time delays of
  the six quasars.  The top and bottom rows show \tab\ and
  \tad\, respectively, plotted against the other relative
  time delays. Colors and symbols are the same as in
  Figure~\ref{fig.parameters}. 
We see correlation between all the relative time
  delays except for \tab. Thus the observational measurements of
  \tab\ and  \tac\ (Dahle et al. 2015) have strong
  constraining power over the model. Furthermore, observational
  measurement of either of the A-D,E,F time delays will narrow the
  uncertainty on the predicted time delays of the other images.}
\label{fig.dtdt}
\end{figure*}

We find strong correlation between the predicted time delays 
 $\tau_{\rm AD}$, $\tau_{\rm AE}$, and $\tau_{\rm AF}$,  as can be
 seen in Figure~\ref{fig.dtdt}.  
Interestingly, $\tau_{\rm AD}$, $\tau_{\rm AE}$, and $\tau_{\rm AF}$
do not correlate with \tab, but they have strong
correlation with \tac. This correlation places a tight
constraint on the predicted time delays of the three central 
images. Moreover, the arrival times of images D, E, and F are strongly
correlated, which means that a time delay measurement of one of them
will provide an additional strict constraint on the time delays of the
other images. 

The correlation of the time delays of the central images is not
surprising. The arrival time lag of the central images is dominated by
gravitational time delay as the light travels close to the center of
mass, due to the deep potential well of the cluster; light will take
longer to travel on this path, although this path is geometrically shorter
(with smaller impact parameter and smaller deflection).
Thus \tac\ is linked to \tad, \tae, \taf\  through its correlation with the overall
normalization of the cluster halo, i.e., the effective velocity
dispersion, $\sigma_0$.

Applying the time delay observational cut on the parameter space, we
are able to narrow down the uncertainties on the predicted time delays
of the central images. Interestingly, we find that these time delays
are short enough to be measured within the next few years: 
  $\tau_{\rm AD}=$\timeADpred, $\tau_{\rm AE}=$\timeAEpred, and
  $\tau_{\rm AF}=$\timeAFpred\ days; 
Moreover, the arrival time of E and F relative to D is
short -- of order 3-5 months: $\tau_{DE}=$\timeDEpred,
$\tau_{DF}=$\timeDFpred\ days, thus measuring $\tau_{\rm DE}$ and $\tau_{\rm
  DF}$ can be achieved within a year or two of cadenced imaging with a
large telescope (Section~\ref{s.conclusions}).

In the following sections, the results of the lensing analysis take into account the 
constraints from the observed time delays, as measured by 
Dahle et al. (2015), and their 95\% confidence interval as described
above. 

\subsection{Cluster Mass}\label{s.mass}
We report the lensing-inferred total projected mass density of the lens (cylindrical
mass) 
 within projected radii of 100, 200, and 500 pc:  
 \monehundred, \mtwohundred, and \mfivehundred, $\pm10\%$ systematic uncertainty. 
The statistical uncertainties are
derived from the MCMC sampling of the parameter space, combined with
the Dahle et al. (2015) 
95\% confidence interval of the time delay measurements (see Section~\ref{s.posterior}).  An additional
10\% systematic uncertainty should be applied, given
the relatively small number of constraints and spectroscopic
redshifts, that limit the
accuracy of the lens model. Johnson \& Sharon (2016) found that while the
enclosed mass is well constrained at the radius of the lensing
evidence, its systematic uncertainty decreases with increasing number of lensing constraints and
spectroscopic redshifts. The analysis in Johnson \& Sharon (2016) is
tuned to the typical number of constraints in high cross-section
lensing clusters such as the Frontier Fields (Lotz et al. 2016), and therefore they do
not sample the affect on systematics in a case like \clustername, a
much lower-mass cluster with four multiply-imaged lensed sources and three spectroscopic
redshifts.
We therefore conservatively adopt a $10\%$ systematic
uncertainty on the enclosed mass, which is the typical uncertainty for
a case of five sources and no spectroscopic redshifts. 
 Interestingly, the observational measurement of the
\tac\ time delay places a tight constraint on the total enclosed
mass and is what drives the relatively small statistical uncertainty. 

Figure~\ref{fig.xray} shows the contours of the projected mass density
distribution from the strong lens model, and the X-ray contours from
\Swift\ observations (Section~\ref{s.swift}). We
find that the X-ray emitting gas and the dark matter distribution are
generally aligned, with no significant offset between their
centroids. A more robust measurement of the X-ray distribution will be
enabled with the superior resolution of Chandra observations. 

\subsection{Magnification}\label{s.time}
The magnification map for a source at the quasar redshift,  $z=\zQSO$,  and the magnifications measured
at the position of each image of the quasar, are shown in
Figure~\ref{fig.magnific}b and Table~\ref{tab.magnific},
respectively. The uncertainties are estimated by computing 
magnification maps for a series of lens models sampled from steps the
MCMC that correspond to $1\sigma$ in the parameter space, and the
95\% confidence interval of the time delay measuremens of Dahle et
al. (2015). 
Since quasars are variable sources and are not standard candles, we
cannot compare the absolute predicted lensing magnification with an
observational measurement. Nevertheless, we can compare the
predictions to the {\it relative} magnifications between images A, B, and
C of the quasar, for which time delays have been measured.  Dahle et
al. (2015) find that the light curves of images A, B, and C, can be
matched with time delays of \tab=\timeABobs\ and
\tac=\timeACobs, and magnitude shift of $\Delta m_{\rm
  AB}=0.340\pm0.007$ mag and $\Delta m_{\rm AC}=0.483\pm0.012$ mag. 
 We find that the model is in agreement with the observed relative
 magnification of image A and B. The model-predicted magnification of
 C is $\sim30\%$ too high to agree with the observed magnification
 ratio between A and C, indicating that the systematic uncertainties
 may be underestimated. 
We note that substructure in the cluster, as well as structure along the line of sight, may contribute to
 discrepancy between the measured and model-derived relative
 magnifications.  

Compared to the initial model in
Dahle et al. (2013), which was based on ground-based observations,
we find that 
the magnifications of A, B, and C are $\sim2.5\times$ higher than
those derived in Dahle et al. (2013), but well within the large statistical uncertainties reported
there. Moreover, systematic uncertainties, which are not taken into
consideration, are large for lens models that are based on few
lensed sources and few spectroscopic redshifts (Johnson \& Sharon
2016). We also note that the new constraints from the \hst\ data
required a more massive component at the south of the cluster
(G4) to explain the lensing evidence that was not identified from the
ground.
Compared to the magnifications in other
wide-separation lensed quasars, we find that the magnifications of A,
B, and C in \clustername\ are similar to the best-fit model-predicted magnifications of the three brightest images in 
SDSSJ1029, from Oguri et al. (2013), while in Oguri et al. (2010), the
lens model of SDSSJ1004 predicts magnifications a factor $\sim2\times$
higher.

\begin{deluxetable}{lclll} 
 \tablecolumns{5} 
\tablecaption{Model-predicted magnifications and time delays
 \label{tab.magnific}} 
\tablehead{
\colhead{Image }   & 
\colhead{F435W} &
\colhead{Magnification} &
\colhead{} &
\colhead{Time delay}  \\
   \colhead{ }   & 
\colhead{magnitude} &
\colhead{$\mu$} &
\colhead{} &
\colhead{[days]}  
} 
\startdata 
A  &21.861& \magA & \nodata  & \nodata \\
B  &22.261& \magB & $\tau_{\rm AB}$     & [~\timeABobs] \\ 
C  &22.227& \magC & $\tau_{\rm AC}$    &[\timeACobs] \\ 
D  &23.827& \magD & $\tau_{\rm AD}$    & \timeADpred \\ 
E  &24.070& \magE & $\tau_{\rm AE}$      &\timeAEpred \\ 
F  &24.909& \magF & $\tau_{\rm AF}$     & \timeAFpred \\ 
\enddata 
 \tablecomments{Magnitudes in the ACS/F435W filter are measured 
within  an aperture of radius $0\farcs56$
in an  observation starting on JD 2456941.06751. 
Time delay is given in days, relative to image A. \tab\
 and \tac\ are observational constraints from Dahle et al. (2015). The
uncertainties represent the 95\% confidence level from the combined
MCMC analysis and the observational time delay constraints.}
\end{deluxetable} 

\begin{figure*}
\centering
\includegraphics[scale=0.4]{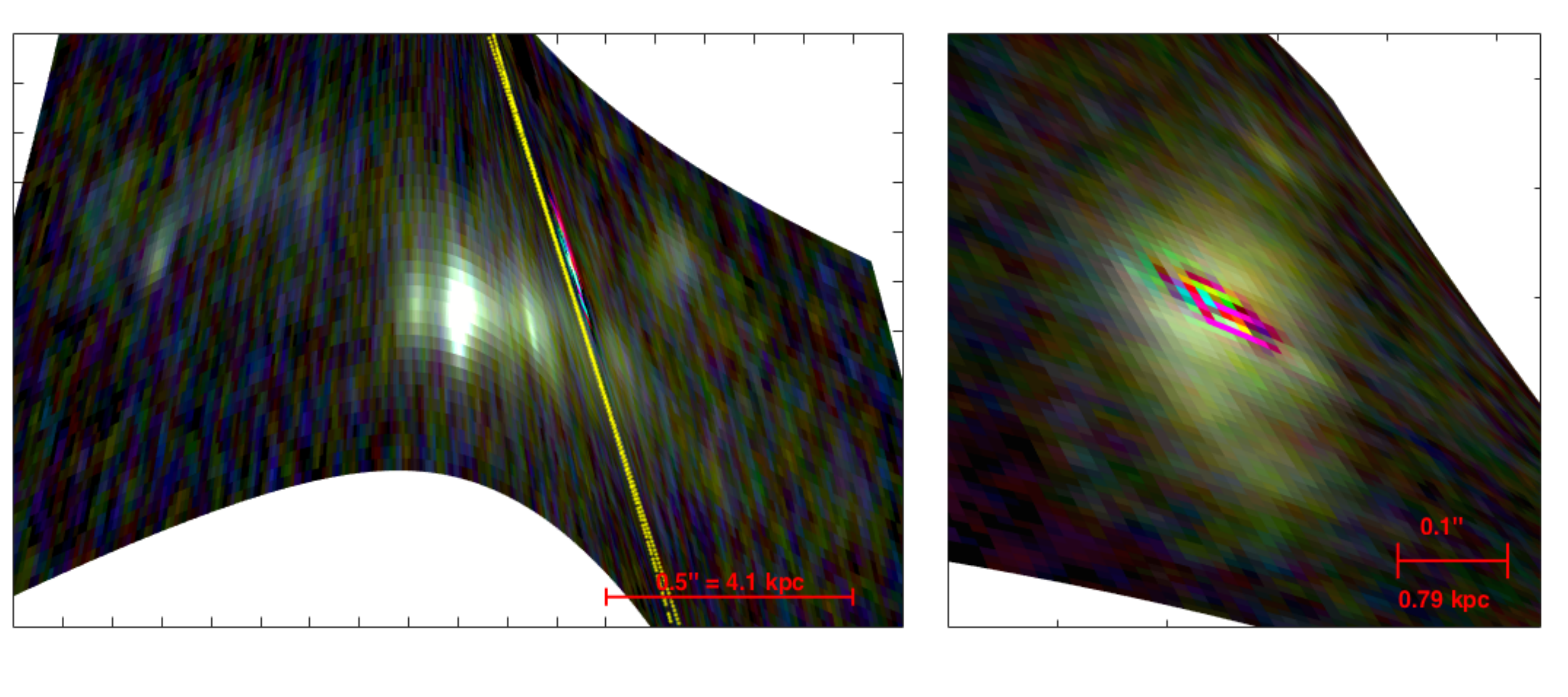}
\caption{ Source plane reconstruction of galaxy A1 at $z=$\zarcA\
  (left) and of the quasar host galaxy at $z=$\zsource\ (right). The
  quasar host is generated from image A of the quasar. The foreground white dwarf and the point-source
  emission from the quasar are masked to reveal the underlying
  information (see text). A
  horizontal bar indicates the scale in arcseconds and kpc at each
  source redshift. The yellow lines in the left panel are the locations
  of the source plane caustics, which map to the critical curves in
  the image plane. }
\label{fig.arcs_src}
\end{figure*}

\subsection{Source Plane Reconstruction}\label{s.source}
We reconstruct the source image of the lensed galaxy A1 at $z=$\zarcA,
and the host galaxy of the quasar at $z=$\zsource, by ray-tracing the
image-plane pixels through the lens equation, $\vec\beta = \vec\theta
-\vec\alpha(\vec\theta)$, where $\vec\beta$ is the source position of each pixel, 
$\vec\theta$ is its observed position, and $\vec\alpha(\vec\theta )$ is the 
deflection matrix scaled by $d_{LS}/d_{S}$, the ratio between the
distance from the lens to the source and from the observer to the
source.  The high lensing magnification resolves small substructure in
these galaxies, which would otherwise be too small for \hst\ resolution. Galaxy A1 is
highly distorted by the lensing potential due to its close proximity
to the caustic.  It is likely that a small region of this galaxy is
multiply imaged within the giant arc.

Prior to ray tracing the images, we subtract the
light of the point source quasar light and the foreground white
dwarf to reveal the underlying information. 
In each band we select a star in the field of view with similar
brightness. We generate a second image by shifting the data so that
the star is at the exact pixel position of the point source we wish to
mask. We then scale the shifted image and subtract it from our data. 
Figure~\ref{fig.arcs_src} shows the reconstructed source-plane image of
galaxy A, and of the quasar host galaxy. 

From the reconstructed source image, A1 measures $\sim13$~kpc in diameter
and the quasar host is measured to be $\sim3$~kpc in diameter. 
A thorough investigation of the physical properties of these galaxies
is left for future work. 

\subsection{Absorbing Systems}\label{s.absorber}
Stark et al. (2013) find strong evidence for an absorption system at
$z=$\zarcA\ in the spectrum of image A of the quasar, indicating that
the extended gas halo of galaxy A1 has neutral hydrogen and metals,
from absorption lines of \Lya, Si~II~$\lambda1526$ and
CIV~$\lambda1549$. Stark et al. (2013) estimate the projected distance
between A1 and image A of the quasar at $\sim50$ kpc.

\begin{figure*}
\centering
\includegraphics[scale=0.5]{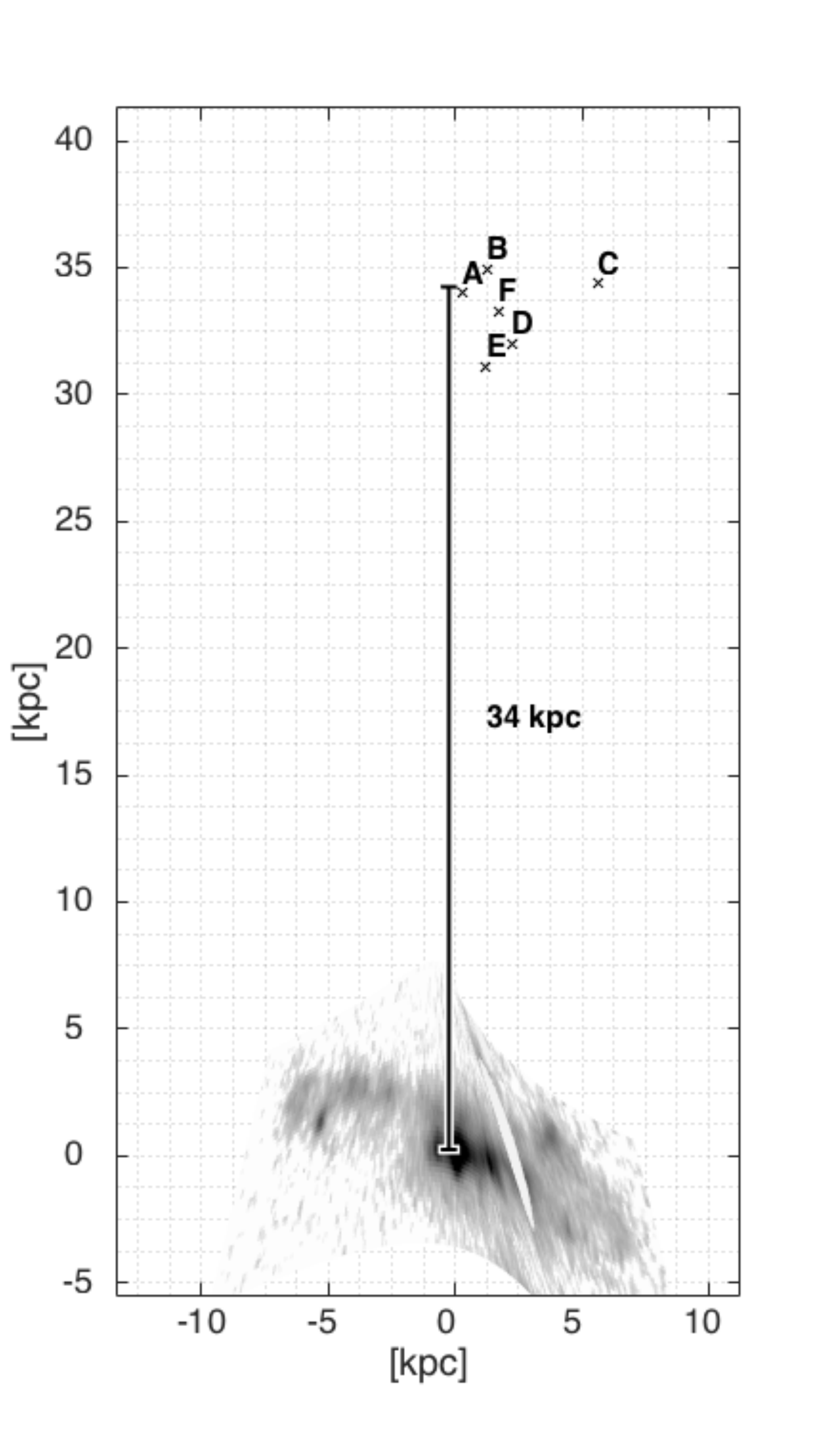}
\includegraphics[scale=0.53]{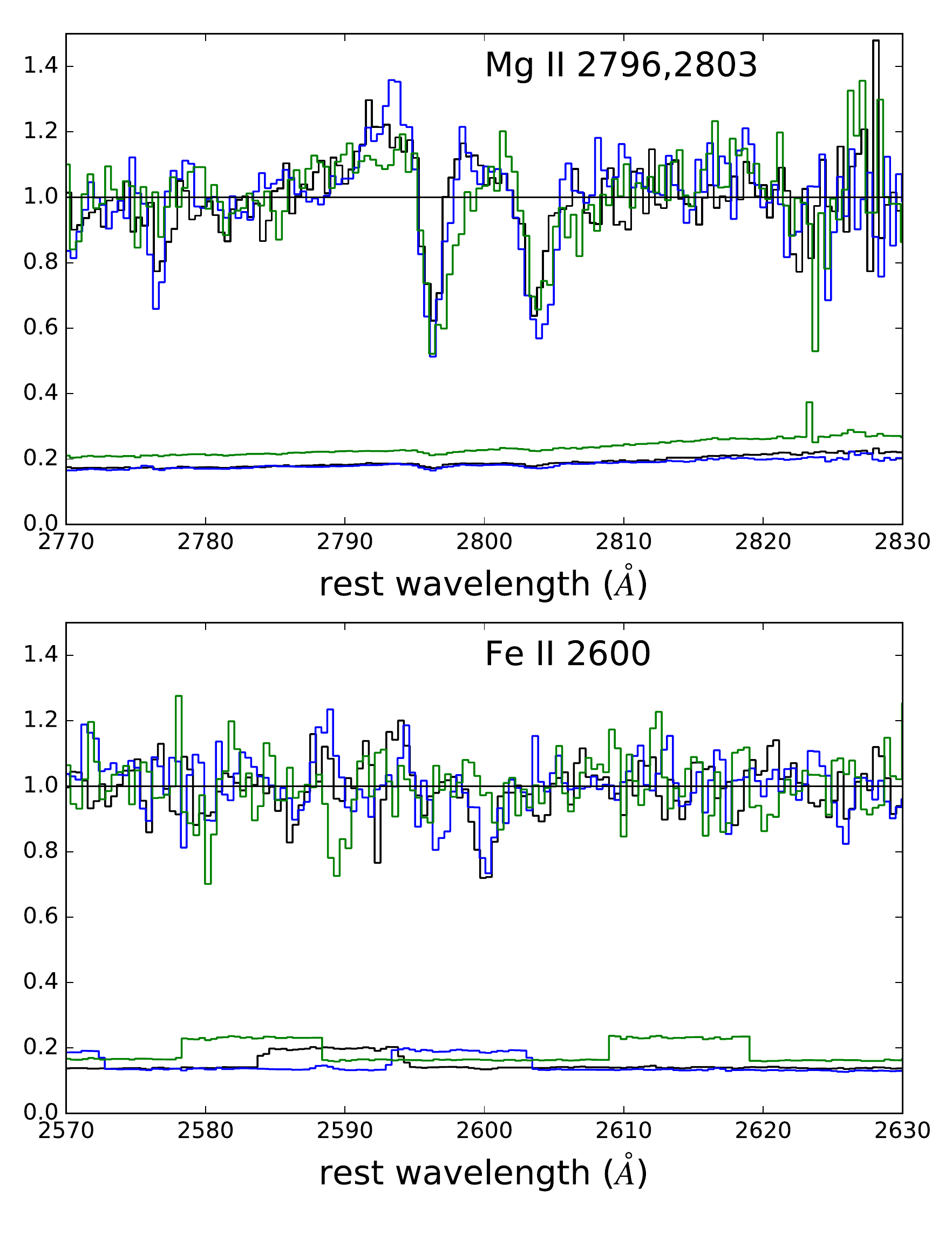}
\caption{{\it Left:} The $z=2.3$ plane, reconstructed from our lens model. At
  this redshift plane, the light from the $z=$\zsource\ quasar passes
  $\sim34$ kpc from galaxy A; the separation between the light rays of
  the quasar
  is a few kpc. The extended gas halo around galaxy A is seen in
  absorption in the spectra of the quasar, which permits a study of
  the spatial distribution of the physical properties of the gas. The
  grid is given in kpc, centered on the brightness peak of galaxy A.
{\it Right: } Continuum-normalized spectra of the intervening absorption system at $z = 2.2988 \pm 0.0002$, for quasar spectra
A (plotted in black), B (blue), and C (green).  The Mg~II~2796, 2803
doublet and the Fe~II~2600 transition are shown at the top and bottom
panels, respectively.
The intervening absorber is clearly detected in Mg~II in all three
quasar spectra, and in Fe~II in quasar images A and B.
}
\label{fig.interloper}
\end{figure*}

A proper estimate of the impact parameter takes into account the
path of the light from the quasar source plane to each of its images,
and where these paths traverse the source plane of A1, at $z=$\zarcA. 
In the left panel of Figure~\ref{fig.interloper} we show a
reconstruction of the source plane at the redshift of galaxy A1. 
By ray-tracing the quasar images to the same redshift of A1, we find
that the quasar light passes \qsoAdist\ kpc north of the center of A1. At
this redshift, the quasar paths are separated by as much as 5 kpc. 
We are therefore presented with a unique opportunity to sample the
uniformity of the gas halo on scales of a few kpc, with at least three bright
lines of sight. 

Our GMOS multi-object spectroscopy masks targeted images A, B, C, and
D of the quasar. Slits were placed on these sources on both nod and
shuffle positions, and on both masks, resulting in a total of 2400 s
on target for A, B, D, and 3600 s on target for C. 
The wavelength coverage allows the detection of FeII~2586,2600 and
MgII~2796,2803 at the redshift of A1.  

The intervening absorption system is detected in the spectra of all
three bright images of the quasar (A, B, C), at a
redshift of $z = 2.2988 \pm 0.0002$.  In the right panel of
Figure~\ref{fig.interloper}, we plot the two strongest features of
this system, 
the  Mg~II~2796,2803~\AA\ doublet and the Fe~II~2600~\AA\ line.  The
spectra were continuum normalized, with the 
continuum calculated by smoothing the spectra with a 40~\AA\ boxcar.  

In Table~\ref{tab.obslog} we tabulate the equivalent width and
redshift measurements for this absorption system in each quasar
spectrum.  Since the blue wing of the Mg~II~2976~\AA\ feature is
affected by the [Ne IV] and Fe~III emission complex at 2423~\AA\
rest-frame \citep{VandenBerk:2001cd}, we do not try to fit this line,
but instead consider the weaker transition Mg~II~2803. 

The intervening system is clearly detected in Mg~II in all three
images of the quasar; the weaker Fe~II~2600 is detected in quasar
images A and B. The equivalent widths are comparable given the
uncertainties listed in Table~\ref{tab.obslog}.  While
Figure~\ref{fig.interloper} shows some variation in the absorption
profiles from quasar image to image, particularly in the amount of
redshifted absorption, these variations may not be significant given the
signal-to-noise ratio of the data.  Deeper spectra are required to
measured differences in the absorption along these three lines of
sight. 

We also detect FeII and MgII absorption from a second absorber at
z=1.202 in the three spectra. The corresponding object is
not currently identified in the imaging or spectroscopic data. 
 The largest separation between the quasar lines
of sight at this redshift is $\sim$40 kpc. 

Co-adding the spectra from the forthcoming spectroscopic followup
campaign (Section~\ref{s.future}) will result in a deep spectrum of each of the quasar images,
and high enough signal to noise to determine some of the physical
properties of the gas halo in the absorbing systems.

\section{Future Work}\label{s.conclusions}\label{s.future}
Ongoing monitoring with the Nordic Optical Telescope (PI: Dahle)
will tighten the constraints on the measured time delays. Recent
observations indicate that image C of the quasar continues its
brightening trend, and the measured 722 day lag provides a unique
opportunity to plan future observations of A and B when they are at
their brightest epoch. During the 2018 observing season, images A and
B will both reach a level $>$1.1 magnitudes brighter than during the
GMOS spectroscopic observations reported in this paper.   
Imaging monitoring with Gemini North  (GN-2016A-Q-28; PI: Gladders) is under
way to constrain the time delays between the internal three images (D,
E, F) of
\clustername. Spectroscopic monitoring with Gemini North (GN-2016B-Q-28; PI: Treu) will enable a
measurement of the mass of the central black hole through
reverberation mapping (Blandford \& McKee 1982; Peterson 1993; Pancoast, Brewer \& Treu 2011). In this paper, we present a revised
lens model of \clustername\ from high resolution multiband \hst\
imaging, new spectroscopic redshifts, and constraints from the
measured time delays of three images of the background quasar.  The
astrophysical applications of \clustername\ span from studies of galaxy
structure at small physical scales, quasar physics,  cluster
astrophysics and cosmology; its investigation has just begun.

\begin{deluxetable}{lll}
\tablecolumns{3}
\tablecaption{Equivalent width and redshift measurements for the
  intervening absorption system 
\label{tab.obslog}}
\tablehead{
\colhead{spectrum} & \colhead{EW\_r(\AA) $\pm \sigma$} & \colhead{redshift}}
\startdata
\cutinhead{Mg II 2803~\AA}
QSO-A  &    $-0.14 \pm 0.09$   &   2.2986\\
QSO-B   &    $-0.26 \pm 0.09$  &    2.2990\\
QSO-C   &   $-0.23 \pm 0.1$    &    2.2989\\
mean      &  $-0.21 \pm 0.05$   &   2.2988\\
\cutinhead{Fe II 2600~\AA}
QSO-A   &   $-0.14 \pm 0.06$  &  2.2981\\
QSO-B    &   $-0.13 \pm 0.08$ &  2.2987\\
QSO-C    &  $-0.04 \pm 0.07$ &   2.2998\\
mean       & $ -0.10 \pm 0.05$ &  2.2989\\
\enddata
\tablecomments{Measured equivalent widths and redshifts for the strongest absorption 
lines in the intervening absorber.  Equivalent widths are in the rest-frame, in \AA, 
and are determined by direct summation. 
Redshifts are taken from the most absorbed pixel for individual quasar spectra. 
For each absorption line, we quote the mean of the measurements across all three 
quasar spectra.}
\end{deluxetable}

\acknowledgments
Support for program GO-13337 was provided by NASA through a grant from the Space Telescope Science Institute, which is operated by the
Association of Universities for Research in Astronomy, Inc., under NASA contract NAS 5-26555.
KEW gratefully acknowledge support by NASA through Hubble Fellowship
grant \#HF2-51368 awarded by the Space Telescope Science Institute, which is operated by the Association of Universities for Research in Astronomy, Inc., for NASA.
Based on observations obtained at the Gemini Observatory, which is operated by the 
Association of Universities for Research in Astronomy, Inc., under a cooperative agreement 
with the NSF on behalf of the Gemini partnership: the National Science Foundation 
(United States), the National Research Council (Canada), CONICYT (Chile), the Australian 
Research Council (Australia), Minist\'{e}rio da Ci\^{e}ncia, Tecnologia e Inova\c{c}\~{a}o 
(Brazil) and Ministerio de Ciencia, Tecnolog\'{i}a e Innovaci\'{o}n
Productiva (Argentina).
This work made use of data supplied by the UK \Swift\ Science Data Centre at the University of Leicester.
This work makes use of the Matlab Astronomy Package (Ofek 2014).

\end{document}

%% file: fixed_parameters.tex
\newcommand{\PcutA}{[$ 1500 $]}

\newcommand{\PxB}{[$0.029 $]}
\newcommand{\PyB}{[$0.036 $]}
\newcommand{\PeB}{[$0.062$]}
\newcommand{\PthetaB}{[$24.3$]}

\newcommand{\PxC}{[$0 $]}
\newcommand{\PyC}{[$0 $]}
\newcommand{\PeC}{[$0 $]}
\newcommand{\PthetaC}{[$0 $]}
\newcommand{\PrcC}{[$0.001 $]}

\newcommand{\PxD}{[$1.895$]}
\newcommand{\PyD}{[$-2.351$]}
\newcommand{\PeD}{[$0.669 $]}
\newcommand{\PthetaD}{[$-15.6$]}

\newcommand{\PxE}{[$1.895 $]}
\newcommand{\PyE}{[$-2.347 $]}
\newcommand{\PeE}{[$0 $]}
\newcommand{\PthetaE}{[$0 $]}
\newcommand{\PrcE}{[$0.001 $]}

\newcommand{\PxF}{[$0.583 $]}
\newcommand{\PyF}{[$-5.339 $]}
\newcommand{\PeF}{[$0.222$]}
\newcommand{\PthetaF}{[$51.1$]}

\newcommand{\PxG}{[$0.583 $]}
\newcommand{\PyG}{[$-5.339 $]}
\newcommand{\PeG}{[$0 $]}
\newcommand{\PthetaG}{[$0 $]}
\newcommand{\PrcG}{[$0.001 $]}

\newcommand{\PxH}{[$-4.425 $]}
\newcommand{\PyH}{[$-14.789$]}

\newcommand{\PxI}{[$3.35$]}
\newcommand{\PyI}{[$5.605 $]}
\newcommand{\PrcI}{[$0.043$]}
\newcommand{\PcutI}{[$7.213$]}

%% file: parameters.tex
\newcommand{\PxA}{$0.26 \pm0.22$}
\newcommand{\PyA}{$2.53 \pm1.20$}
\newcommand{\PeA}{$0.29 \pm0.05$}
\newcommand{\PthetaA}{$95.8 \pm2.1$}
\newcommand{\PrcA}{$42.7 \pm6.5$}
\newcommand{\PsigmaA}{$668 \pm 22$}
\newcommand{\PrcB}{$0.13 \pm0.13$}
\newcommand{\PsigmaB}{$274 \pm 85$}
\newcommand{\PcutB}{$24.9 \pm23.5$}
\newcommand{\PsigmaC}{$30.3 \pm19.5$}
\newcommand{\PcutC}{$23.6 \pm23.2$}
\newcommand{\PrcD}{$0.15 \pm0.13$}
\newcommand{\PsigmaD}{$294 \pm104$}
\newcommand{\PcutD}{$11.0 \pm8.3$}
\newcommand{\PsigmaE}{$29.2 \pm18.8$}
\newcommand{\PcutE}{$24.9 \pm20.9$}
\newcommand{\PrcF}{$0.16 \pm0.13$}
\newcommand{\PsigmaF}{$209 \pm 28$}
\newcommand{\PcutF}{$27.7 \pm22.3$}
\newcommand{\PsigmaG}{$32.2 \pm17.5$}
\newcommand{\PcutG}{$26.7 \pm22.8$}
\newcommand{\PeH}{$0.49 \pm0.20$}
\newcommand{\PthetaH}{$-74.7 \pm3.9$}
\newcommand{\PrcH}{$3.56 \pm1.32$}
\newcommand{\PsigmaH}{$360 \pm129$}
\newcommand{\PcutH}{$18.5 \pm14.6$}
\newcommand{\PeI}{$0.66 \pm0.13$}
\newcommand{\PthetaI}{$-60.0 \pm9.7$}
\newcommand{\PsigmaI}{$56.2 \pm43.2$}

%% file: spectable.tex
\begin{deluxetable*}{llllll} 
 \tablecolumns{6} 
\tablecaption{Lensing Constraints and Spectroscopy Results
 \label{tab.spectroscopy}} 
\tablehead{
\colhead{ID}   & 
\colhead{R.A.} &
\colhead{Decl.} &
\colhead{Redshift}    & 
\colhead{mask obj. ID}    & 
\colhead{Comments}\\
&
\colhead{[J2000]} &
\colhead{[J2000]} &
&&
} 
\startdata 
QSO-A  &335.537707 & 27.760543  & $2.8050\pm0.0006$& 2-1001, 2-1101 & HeII~1640+OIII]1666+[OII]2470 emission \\
QSO-B  & 335.536690  & 27.761119  & $2.8048\pm  0.0005$ &2-118, 2-8163&HeII~1640+OIII]1666 emission \\
QSO-C  &335.532960  & 27.760505  & $2.8050\pm  0.0006$ &1-8166, 2-1113, & HeII~1640+CII]2327+[OII]2470 emission \\
& & & &  2-8179 & \\
QSO-D       &335.536205 & 27.758901  & $2.8012\pm  0.0005$ &2-1104, 2-1114 & Dominated by light from G2; redshift from CIV+CIII]  \\
QSO-E       &335.536007 & 27.758248  & \nodata &\nodata & No new data; spec confirmed by Dahle et al. (2013)  \\
QSO-F       &335.535874 & 27.759723  & \nodata &\nodata & No new data; spec confirmed by Dahle et al. (2013)  \\
QSO-host-A  &335.537968 & 27.760220  & $2.8$ &1-1502, 1-1512 &Low S/N; Dominated by quasar spectrum \\
A1 &335.536022 &27.756889  & $2.2963\pm0.0004$ &2-137, 2-8180 &NIII] 1750, SiIII] 1892, CIII] 1909 Nebular emission\\
A1 &335.536909 & 27.756990& \nodata & 1-1501, 1-1511& Faint end; No signal\\
B1  &335.53388 & 27.757979  & $4.562\pm0.002$ &1-1202, 1-1211&Shapley composite comparison;  \Lya\ at z=4.5651, \\
&&&&2-1201, 2-1211&HeII~1640 at z=4.5564 \\
B2  &335.534820 &  27.757630  & \nodata &1-1212, 1-1213 &low S/N or contaminated \\
B3  &335.538410 &  27.758236  & \nodata &2-1201, 2-1213 &low S/N or contaminated \\ 
C1 & 335.533620 &27.760879   & \nodata &\nodata & \nodata  \\
C2  &335.538420 & 27.760385  &\nodata &1-1311 & low S/N (see text) \\
C3  &335.538425 & 27.760429  &\nodata &1-1303, 1-1313& low S/N (see text) \\ 
D1  &335.533530 &  27.755175  & ($0.837\pm 0.002$) &1-1411, 1-1401&Uncertain redshift; probably contaminated by FG object. \\
D2  &335.534090& 27.754942   & ($0.836 \pm0.001$) &2-1402, 2-1412 &Uncertain redshift; probably contaminated by FG object.  \\
D3  &335.534540 & 27.754882  & \nodata &1-1412, 1-1402 &Sky position contaminated \\
\hline
cluster gal  G1 &335.535793 &  27.759830  & $0.4901\pm 0.0002$ &1-112, 2-8168 & cluster galaxy \\
cluster gal  G2 & 335.536366 & 27.759190 & $0.4925 \pm 0.0002$ & 1-8155              & cluster galaxy, $0\farcs5$ slit\\
cluster gal  G3 & 335.536022 & 27.758369 & $0.4919 \pm 0.0004$ & 2-135, 2-8177 & cluster galaxy, z=2.8055 CIV emission from QSO-E\\
cluster gal  G4 & 335.534391 & 27.755760 & $0.4922 \pm 0.0002$ & 2-148               & cluster galaxy\\
cluster gal & 335.525723 & 27.738350 & $0.4906 \pm 0.0008$ & 2-232 &cluster galaxy\\
cluster gal & 335.527496 & 27.751221 & $0.4902 \pm 0.0004$ & 2-186 & cluster galaxy\\
cluster gal & 335.533733 & 27.753309 & $0.4861 \pm 0.0003$ & 1-8170, 2-163  & cluster galaxy\\
cluster gal & 335.536966 & 27.744699 & $0.4945 \pm 0.0006$ & 2-176 & cluster galaxy\\
cluster gal & 335.553675 & 27.773190 & $0.4913 \pm 0.0002$ & 2-8104 & cluster galaxy\\
cluster gal &  335.535707 &  27.755211 & $0.4893 \pm 0.0007$ & 2-8186& cluster galaxy\\
cluster gal &  335.535421&  27.754869 &  $0.4883 \pm 0.0006$ & 2-8189& cluster galaxy\\
\hline
BG & 335.517597 & 27.782749 & $0.732 \pm 0.001$ & 2-8155 & background\\
BG & 335.546093 & 27.751680 & $0.6031 \pm 0.0002$ & 1-8145, 2-119 & background, strong nebular emission, likely NL AGN\\
BG & 335.516653 & 27.766451 & $0.8813 \pm 0.0001$ & 2-169 & background, star foming; strong nebular emission\\
\hline
FG & 335.505838 & 27.762159 & $0.3490 \pm 0.0001$ & 1-194 & foreground, strong H$\alpha$\\
FG & 335.520544 & 27.755730 & $0.4487 \pm 0.0002$ & 2-196 & foreground\\
\hline
cluster gal & 335.546740 & 27.758008 & 0.4833 $\pm$ 0.0001 & \nodata & SDSS-DR9\\ 
cluster gal & 335.535580 & 27.772658 & 0.4883 $\pm$ 0.0001 & \nodata & SDSS-DR9\\ 
cluster gal & 335.503210 & 27.789785 & 0.4843 $\pm$ 0.0001 & \nodata & SDSS-DR9
\enddata 
 \tablecomments{Spectroscopic redshifts, from Gemini/GMOS
   observations, of images of the quasar, lensed galaxies, cluster
   member, foreground and background galaxies. The coordinates of the six images of the quasar, as well as galaxies B1-3, C1-3, and D1-3 correspond to the 
exact coordinate of their peak brightness, that was used as lensing constraint. Otherwise, the coordinates on which the slits were placed are given. Due to the uncertain spectroscopy result for arc C and D, we left their redshifts as free parameters with a priors set by photometric redshift analysis, $2.0\le z_C\le4.0$ and $3.8\le z_D\le 5.0$. All other redshifts were fixed to their spectroscopic measurements.
 Stars and slits with
   insufficient data quality are not shown. See also Figure~\ref{fig.masks}.}
\end{deluxetable*}